\documentclass{sbc2019}%

\usepackage[utf8]{inputenc}
\usepackage[misc,geometry]{ifsym} 
\usepackage{fontawesome}
\usepackage{academicons}
\usepackage{hyperref}
\usepackage{aas_macros}
\usepackage[bottom]{footmisc}
\usepackage{supertabular}
\usepackage{afterpage}
\usepackage{pifont}
\usepackage{multicol}
\usepackage{multirow}

\usepackage{algorithm}
\usepackage{algpseudocode}

\usepackage{colortbl} %
\usepackage{hhline}

\usepackage{xspace}
\usepackage{graphicx}
\usepackage[caption=false]{subfig}
\usepackage{relsize}
\usepackage{stmaryrd}

\usepackage{datetime}
\usepackage{url}
\usepackage{rotating}
\usepackage{array}

\usepackage{ragged2e}
\usepackage{scalerel}
\usepackage[acronym,nohypertypes={acronym}]{glossaries}

\usepackage{booktabs, tabularx}

\usepackage{fancyvrb} %

\setcitestyle{square}

\definecolor{orcidlogo}{rgb}{0.37,0.48,0.13}
\definecolor{unilogo}{rgb}{0.16, 0.26, 0.58}
\definecolor{maillogo}{rgb}{0.58, 0.16, 0.26}
\definecolor{darkblue}{rgb}{0.0,0.0,0.0}
\hypersetup{colorlinks,breaklinks,
            linkcolor=darkblue,urlcolor=darkblue,
            anchorcolor=darkblue,citecolor=darkblue}

\jid{JBCS}
\jtitle{Journal of Internet Services and Applications, 2025, 16:2, }
\doi{10.5753/jisa.2025.XXXXXX}
\copyrightstatement{This work is licensed under a Creative Commons Attribution 4.0 International License}
\jyear{2025}

\usepackage{soulutf8}
\definecolor{green}{RGB}{0,150,0}

\newcommand{\serie}{\ensuremath{\mathcal{P}}\xspace}
\newcommand{\couple}{p} 
\newcommand{\abcisse}{t} 
\newcommand{\ordonnee}{x} 
\newcommand{\tqueried}{\abcisse}

\newcommand{\datastructure}{\ensuremath{\mathcal{F}}\xspace}
\newcommand{\history}{\ensuremath{\mathcal{H}}\xspace}

\newcommand{\segment}{\ensuremath{\vec{s}}\xspace}
\newcommand{\scurr}{\ensuremath{\vec{s}_M}\xspace}
\newcommand{\gradient}{\ensuremath{g}\xspace} 
\newcommand{\acurr}{\ensuremath{\gradient_M}\xspace}
\newcommand{\amin}{\ensuremath{\gradient_{\text{min}}}\xspace}
\newcommand{\amax}{\ensuremath{\gradient_{\text{max}}}\xspace}

\newcommand{\pcurr}{\ensuremath{\history\left[M\right]}\xspace}
\newcommand{\tcurr}{\abcisse_M}
\newcommand{\vcurr}{\ordonnee_M}

\newcommand{\plast}{\ensuremath{\couple_\text{last}}\xspace}
\newcommand{\tlast}{\abcisse_\text{last}}
\newcommand{\vlast}{\ordonnee_\text{last}}
\newcommand{\pinserted}{\ensuremath{\couple}\xspace}
\newcommand{\tinserted}{\abcisse}
\newcommand{\vinserted}{\ordonnee}
\newcommand{\ainserted}{\ensuremath{\gradient}\xspace}

\newcommand{\tnorm}{\abcisse_\Delta}
\newcommand{\vnorm}{\ordonnee_\Delta}

\makeatletter
\newcommand*{\shifttext}[2]{%
  \settowidth{\@tempdima}{#2}%
  \makebox[\@tempdima]{\hspace*{#1}#2}%
}
\makeatother

\definecolor{Gray}{gray}{0.88}

\newacronym[longplural=\textit{Location Privacy Protection Mechanisms},shortplural=LPPMs]{lppm}{LPPM}{\textit{Location Privacy Protection Mechanism}}
\newacronym[longplural=\textit{Points of Interest},shortplural=POIs]{poi}{POI}{\textit{Point Of Interest}}
\newacronym{iot}{IoT}{\textit{Internet of Things}}
\newacronym{spi}{SPI}{\textit{Sensitive Personal Information}}
\newacronym{gps}{GPS}{\textit{Global Positioning System}}
\newacronym{cdf}{CDF}{\textit{Cumulative Distribution Function}}
\newacronym{pla}{PLA}{\textit{Piece-wise Linear Approximation}}
\newacronym{flair}{FLI}{\textit{Fast Linear Interpolation}}
\newacronym{intact}{INTACT}{\textit{IN-siTu locAtion proteCTion}}
\newacronym{ds}{D\&S}{\textit{Divide\,\&\,Stay}}
\newacronym{lbs}{LBS}{\textit{Location-Based Service}}
\newacronym{adb}{ADB}{\textit{Android Debug Bridge}}

\newcommand{\FLI}{\acrshort{flair}\xspace}
\newcommand{\intact}{\acrlong{intact}\xspace}
\newcommand{\INTACT}{\acrshort{intact}\xspace}
\newcommand{\stay}{\acrlong{ds}\xspace}
\newcommand{\STAY}{\acrshort{ds}\xspace}
\newcommand{\poiattack}{{\sc \gls{poi}-Attack}\xspace}

\usepackage{color}

\begin{document}

\title[In-situ location protection]{\INTACT: Compact~Storage of Data~Streams in Mobile~Devices to Unlock User Privacy at the Edge}

\author[Raes et al. 2025]{
\affil{\textbf{Rémy Raes}~\href{https://orcid.org/0000-0002-8239-7683}{\textcolor{orcidlogo}{\aiOrcid}}~\textcolor{blue}{\faEnvelopeO}~~[~\textbf{Inria, Univ.\,Lille, CNRS, UMR\,9189 CRIStAL, France}~|\href{mailto:remy.raes@inria.fr}{~\textbf{\textit{remy.raes@inria.fr}}}~]}

\affil{\textbf{Olivier Ruas}~\href{https://orcid.org/0000-0002-6862-9046}{\textcolor{orcidlogo}{\aiOrcid}}~~[~\textbf{Pathway, France}~|\href{mailto:olivier.ruas@gmail.com}{~\textbf{\textit{olivier.ruas@gmail.com}}}~]}

\affil{\textbf{Adrien Luxey-Bitri}~\href{https://orcid.org/0000-0003-1777-307X}{\textcolor{orcidlogo}{\aiOrcid}}~~[~\textbf{Inria, Univ.\,Lille, CNRS, UMR\,9189 CRIStAL, France}~|\href{mailto:adrien.luxey@inria.fr}{~\textbf{\textit{adrien.luxey@inria.fr}}}~]}

\affil{\textbf{Romain Rouvoy}~\href{https://orcid.org/0000-0003-1771-8791}{\textcolor{orcidlogo}{\aiOrcid}}~~[~\textbf{Inria, Univ.\,Lille, CNRS, UMR\,9189 CRIStAL, France}|\href{mailto:romain.rouvoy@inria.fr}{~\textbf{\textit{romain.rouvoy@inria.fr}}}~]}
}

\begin{frontmatter}
  \maketitle

\begin{mail}
  Centre Inria de l'Université de Lille, Parc scientifique de la Haute-Borne, 40, avenue Halley - Bât A - Park Plaza, 59650 Villeneuve d'Ascq - France
\end{mail}

\begin{dates}
  \small{\textbf{Received:} 29 Nov 2024~~~$\bullet$~~~\textbf{Accepted:} 12 May 2025~~~$\bullet$~~~\textbf{Published:} DD Month YYYY}
\end{dates}

  \begin{abstract}

    Data streams produced by mobile devices, such as smartphones, offer highly valuable sources of information to build ubiquitous services.
    Such data streams are generally uploaded and centralized to be processed by third parties, potentially exposing sensitive personal information.
    In this context, existing protection mechanisms, such as \glspl{lppm}, have been investigated.
    Alas, none of them have actually been implemented, nor deployed in real-life, in mobile devices to enforce user privacy at the edge.
    Moreover, the diversity of embedded sensors and the resulting data deluge makes it impractical to provision such services directly on mobiles, due to their constrained storage capacity, communication bandwidth and processing power.
    
    This article reports on the \FLI technique, which leverages a piece-wise linear approximation technique to capture compact representations of data streams in mobile devices.
    Beyond the \FLI storage layer, we introduce \stay, a new privacy preservation technique to execute \glspl{poi} inference.
    Finally, we deploy both of them on Android and iOS as the \INTACT framework, making a concrete step towards enforcing privacy and trust in ubiquitous computing systems.%
  \end{abstract}

\end{frontmatter}

  \section{Introduction}\label{sec:intro}
\newcommand{\GPSsecond}{2\xspace}
\newcommand{\Accsecond}{476\xspace}
\newcommand{\GPSday}{172{,}800\xspace}
\newcommand{\Accday}{41{,}126{,}400\xspace}
\newcommand{\scaleToyExample}{0.63}

\begin{figure}[t]
    \centering
    \subfloat[Cabspotting mobility sub-trace of \texttt{user~0}.]{
      \label{fig:cabspotting_subtrace}
      \includegraphics[width=0.9\linewidth]{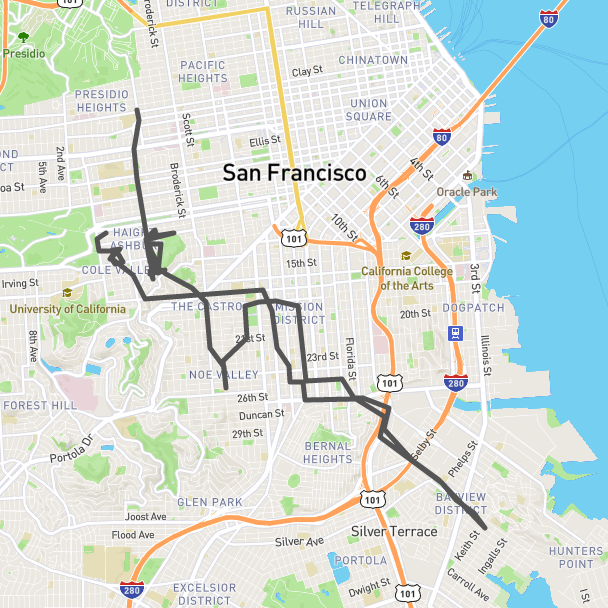}
    }
    \centering

    \subfloat[Raw longitude trace for \texttt{user~0}.]{
      \label{fig:real_example_lng_raw}
      \includegraphics[width=.46\linewidth]{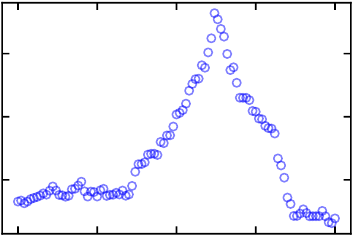}
    }%
    \subfloat[Modeled longitude with \FLI.]{
      \label{fig:real_example_lng_fli}
      \includegraphics[width=.46\linewidth]{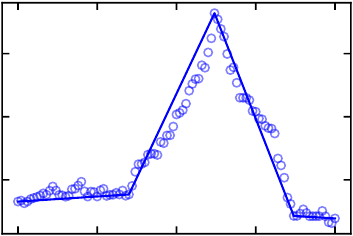}
    }

    \subfloat[Raw latitude trace for \texttt{user~0}.]{
      \label{fig:real_example_lat_raw}
      \includegraphics[width=.46\linewidth]{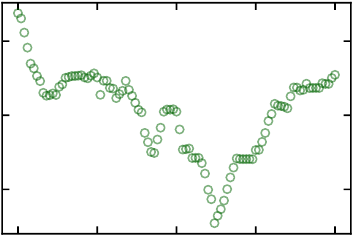}
    }%
    \subfloat[Modeled latitude with \FLI.]{
      \label{fig:real_example_lat_fli}
      \includegraphics[width=.46\linewidth]{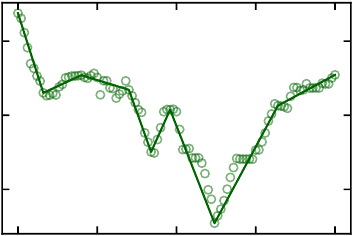}
    }
  \caption{\FLI compacts any location stream as a sequence of segments.}
  \label{fig:real_example}
\end{figure}

\paragraph{Mobile devices: usage and data}
With the advent of smartphones and more generally the \gls{iot}, ubiquitous devices are mainstream in our societies and widely deployed at the edge of networks.
Such constrained devices are not only consuming data and services, such as content streaming, restaurant recommendations or more generally \glspl{lbs}, but are also key producers of data streams by leveraging a wide variety of embedded sensors that capture the surrounding environment of end-users, including their daily routines.
The data deluge generated by a connected user is potentially tremendous: according to preliminary experiments, a smartphone can generate approximately \GPSsecond pairs of \gls{gps} samples and \Accsecond triplets of accelerometer samples per second, resulting in more than \GPSday location and \Accday acceleration samples daily.

In this context, the storage and processing of such data streams in mobile devices are challenges that cannot only be addressed by assuming that the hardware capabilities will keep increasing.
In particular, sustainability issues call for increasing the lifespan of legacy devices, thus postponing their replacement.
This implies that software-defined solutions are required to leverage the shortenings of hardware resources.

\paragraph{Privacy}
Continuous data streams inevitably include \gls{spi} that jeopardize the privacy of end-users, when processed by malicious stakeholders. %
While machine learning algorithms are nowadays widely adopted as a convenient keystone to process large datasets and infer actionable insights, they often require grouping raw input datasets in a remote place, thus imposing a privacy threat for end-users sharing their data.
This highlights the utility vs. privacy trade-off that is inherent to any data-sharing activity~\citep{pulp}.
On the one hand, without crowd-sourced \gls{gps} traces, it would be hard to model traffic in real-time and recommend itineraries. %
On the other hand, it is crucial to protect user privacy when accepting to gather \gls{spi}.

\paragraph{\acrshort{lppm}}
To address this ethical challenge, privacy-preserving machine learning~\citep{xu2015privacy} and decentralized machine learning~\citep{bellet2017fast,y2018decentralized} are revisiting state-of-the-art machine learning algorithms to enforce user privacy, among other properties.
Regarding location privacy, several \acrfullpl{lppm} have been developed to preserve user privacy in mobility situations.
Location reports are evaluated and obfuscated before being sent to a service provider, hence keeping user data privacy under control.
The user no longer automatically shares their data streams with service providers but carefully selects what they share and make sure the data they unveil does not contain any \gls{spi}.
For example, Geo-Indistinguishability~\citep{geoI} generalizes differential privacy~\citep{differential} to \gls{gps} traces, while {\sc Promesse}~\citep{promesse} smooths the \gls{gps} traces---both temporally and geographically---to erase \glspl{poi} from the input trace.
\glspl{lppm} successfully preserve sensitive data, such as \glspl{poi}, while maintaining the data utility for the targeted service.

\paragraph{Challenges}
Despite their reported effectiveness, no \gls{lppm} has ever been implemented and deployed on mobile devices: previous works have been simulated on the \acrfull{adb}~\citep{eden} at best.
While the extension of those works to Android and iOS devices may seem straightforward, it is hindered by the scarce resources of edge devices.
Storage is notably challenging, as \glspl{lppm} generally require the user to access all their \gls{gps} traces, and sometimes the ones of additional users.
To this day, storing such amounts of data is challenging on constrained mobile devices.
Data processing algorithms are additionally not optimized for mobile devices.
Furthermore, even if the storage capacity of modern devices keeps increasing, the deluge of data streams produced by all the sensors of a smartphone (e.g., \gls{gps}, accelerometer, gyroscope, etc.) makes it impossible to store all the raw data for the applications that require it.

\paragraph{Content}

This work, as an extension of the method presented in~\citep{fli}, demonstrates that modeling data streams successfully addresses these device-level storage \& processing challenges.

In previous work from our team, a novel algorithm, \acrfull{flair}, was introduced  to model and store data streams under memory constraints, leveraging a \gls{pla} technique~\citep{fli}.
\FLI does not store raw data samples (Fig.~\ref{fig:real_example_lng_raw}~\&~\ref{fig:real_example_lat_raw}) but, instead, models their evolution as linear interpolations (Fig.~\ref{fig:real_example_lng_fli}~\&~\ref{fig:real_example_lat_fli})---offering a much bigger storage capacity at the cost of a controlled approximation error.

In this paper, we demonstrate how \FLI can be leveraged to implement an \gls{lppm} working directly on mobile phones---thanks to the increased \gls{gps} storage capacity offered by \FLI.
To be proven useful, the \gls{lppm}'s privacy gains must be evaluated \emph{in situ}, before any geolocation trace is shared.
We thus also introduce a new \gls{poi} attack algorithm, dubbed \acrfull{ds}, running directly on mobile, which goal is to extract \glspl{poi} from GPS traces.
We will use \STAY to assert whether the \glspl{lppm} effectively protects the privacy of traces modeled with \FLI.

\begin{figure}
  \center
  \includegraphics[width=.9\linewidth]{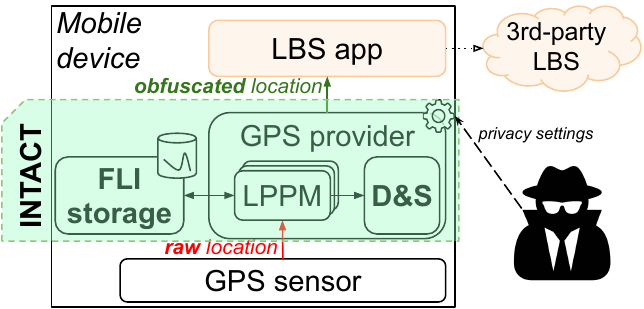}
  \caption{Overview of INTACT contributions.}
  \label{fig:intact}
\end{figure}

Altogether, \FLI and \STAY constitute the \intact (\INTACT) framework, pictured in Fig.~\ref{fig:intact}.
The figure shows how our two contributions can be combined to enable the deployment of \glspl{lppm} on mobile devices to support location-preserving \gls{lbs}.
We report that \INTACT can store vast amounts of location data on mobile, and that it can protect end-user privacy while using \gls{lbs} by using device-local \glspl{lppm}.

In the following, we first discuss related works (Sec.~\ref{sec:related}), before dissecting the \INTACT framework (Sec.~\ref{sec:approach}). 
Next, we present our experimental setup (Sec.~\ref{sec:setup}) and the results we obtained (Sec.~\ref{sec:eval}).
Finally, we discuss the limitations of our approach (Sec.~\ref{sec:threats}) and then conclude this paper (Sec.~\ref{sec:conclusion}).

  \section{Related Work}
\label{sec:related}

\subsection{Data Storage}

Overcoming the memory constraints of mobile devices to store data streams usually implies the integration of efficient temporal databases.
To take the example of Android: few databases are available, such as {\sc SQLite} and its derivative {\sc Drift}~\citep{drift}, the cloud-supported Firebase~\citep{firebase}, the {\sc NoSQL hive}, and {\sc ObjectBox}~\citep{objectbox}.
The situation is similar on iOS.

\emph{Relational databases} (\emph{e.g.}, SQL) are typically designed for \emph{OnLine Transactional Processing} (OLTP) and \emph{OnLine Analytical Processing} (OLAP) workloads, which widely differ from time-series workloads.
In the latter, reads are mostly contiguous (as opposed to the random-read tendency of OLTP); writes are most often inserts (not updates) and typically target the most recent time ranges.
OLAP is designed to store big data workloads to compute analytical statistics, while not putting the emphasis on read or write performances.
Finally, in temporal workloads, it is unlikely to process writes \& reads in the same single transaction~\citep{timescale_building_2019}.

\emph{Time~series databases (TSDB).}
Despite these deep differences, several relational databases offer support for temporal data with industry-ready performance---\emph{e.g.}, {\sc TimescaleDB}~\citep{timescale} is a middleware that exposes temporal functionalities atop a relational {\sc PostgreSQL} foundation.
\textsc{InfluxDB}~\citep{InfluxDB} is one of the most widely used temporal databases.
Unfortunately, when facing memory constraints, its retention policy prevents the storage from scaling in time: the oldest samples are dumped to make room for the new ones.
Furthermore, on mobile, memory shortages often cause the operating system to kill the TSDB process to free the memory, which is opposed to the very concept of in-memory databases.

\emph{Moving objects databases (MOD).}
Location data storage is an issue that has also been studied in the MOD community, where a central authority merges trajectory data from several sensors in real-time.
To optimize storage and communication costs, it does not store the raw location data, but rather trajectory approximations.
\emph{Linear Dead Reckoning} (LDR)~\citep{ldr} limits data exchange between sensors and server by sending new location samples only when a predefined \textit{accuracy bound $\epsilon$} (in meters) is exceeded.
A mobility prediction vector is additionally shared every time a location sample is sent.
Even so, this class of solutions requires temporarily storing modeled locations to ensure they fit the $\epsilon$ bound and exclusively focuses on modeling location data streams, while we aim at storing any type of real-valued stream.

\emph{Modeling data streams.}
While being discrete, the streams sampled by sensors represent inherently continuous signals.
Data modeling does not only allow important memory consumption gains, but also flattens sensors' noise, and enables extrapolation between measurements.
In particular, \acrfull{pla} is used to model the data as successive affine functions.
An intuitive way to do linear approximation is to apply a bottom-up segmentation: each pair of consecutive points is connected by interpolations; the less significant contiguous interpolations are merged, as long as the obtained interpolations introduce no error above a given threshold.
The bottom-up approach has low complexity, but usually requires an offline approach to consider all the points at once.
The \emph{Sliding Window And Bottom-up} (SWAB) algorithm~\citep{swab}, however, is an online approach that uses a sliding window to buffer the latest samples on which a bottom-up approach is applied.
{\sc emSWAB}~\citep{emswab} improves the sliding window by adding several samples at the same time instead of one.
Instead of interpolation, linear regression can also be used to model the samples reported by IoT sensors~\citep{grutzmacher2018time}.
For example, {\sc Greycat}~\citep{greycat} adopts polynomial regressions with higher degrees to further compress the data.
Unfortunately, none of those works have been implemented on mobile devices to date.

\textsc{Sprintz}~\citep{blalockSprintzTimeSeries2018} proposes a mobile lossless compression scheme for multi-modal integer data streams, along with a comparison of other compression algorithms. 
They target streaming of the compressed data to a centralized location from \gls{iot} devices with minimal resources. 
This work is orthogonal to ours, as \FLI intends to model floating-point unimodal streams on one's devices for further local computation, instead of streaming it to a third-party server.

Closer to our work, FSW~\citep{liu2008novel} and the {\sc ShrinkingCone} algorithm~\citep{galakatos2019fiting} attempt to maximize the length of a segment while satisfying a given error threshold, using the same property used in \FLI.
FSW is not a streaming algorithm as it considers the dataset as a whole, and does not support insertion.
The {\sc ShrinkingCone} algorithm is a streaming greedy algorithm designed to approximate an index, mapping keys to positions: it only considers monotonic increasing functions and can produce disjoints segments.
\FLI models non-monotonic functions in a streaming fashion, while providing joint segments.

\textbf{Limitations.}
To the best of our knowledge, state-of-the-art storage solutions for unbounded data streams either require storing raw
data samples or triggering {\em a~posteriori} data computations, which makes them unsuitable for mobile devices.

\subsection{Location Privacy Attacks}
Raw user mobility traces can be exploited to model the users' behavior and reveal their \acrfull{spi}.
In particular, the \glspl{poi} are widely used as a way to extract \gls{spi} from mobility traces.
In a nutshell, a \gls{poi} is a place where the user comes often and stays for a significant amount of time: it can reveal their home, workplace, or leisure habits.
From revealed \glspl{poi}, more subtle information can also be inferred: sexual orientation from attendance to LGBT+ places, for instance.
The set of \glspl{poi} can also be used as a way to re-identify a user in a dataset of mobility traces~\citep{primault2014differentially,gambs2014anonymization}.
The \glspl{poi} can be extracted using spatiotemporal clustering algorithms~\citep{zhou2004discovering,hariharan2004project}.
Alternatively, an attacker may also re-identify a user directly from raw traces, without computing any \gls{poi}~\citep{maouche2017ap}.

\subsection{Protecting Mobility Datasets}
When data samples are gathered in a remote server, one can expect the latter to protect the dataset as a whole.
In particular, {\it k-anonymity}~\citep{sweeney2002k} is the property of a dataset guaranteeing that whenever some data leaks, the owner of each data trace is indistinguishable from at least $k-1$ other users contributing to the dataset.
Similarly, {\it l-diversity}~\citep{machanavajjhala2007diversity} extends {\it k-anonymity} by ensuring that the $l$ users are diverse enough not to infer \gls{spi} about the data owner.
Finally, {\it differential privacy}~\citep{differential} aims at ensuring that the inclusion of a single element in a dataset does not alter significantly an aggregated query on the whole dataset.
However, all these techniques require personal samples to be grouped to enforce user privacy.

\subsection{Protecting Individual Traces}
Rather than protecting the dataset as a whole, each data sample can also be protected individually.
In the case of location data, several protection mechanisms---called \acrfullpl{lppm}---have been developed.
They may be deployed in a remote server where all data samples are gathered, or run directly on the source device before any data exchange.

\paragraph{{\em Geo-Indistinguishability} (\textsc{GeoI})~\citep{geoI}} implements differential privacy~\citep{differential} at the trace granularity.
In particular, {\sc GeoI} adjusts mobility traces with two-dimensional Laplacian noise, making \glspl{poi} more difficult to infer.
{\it Heat Map Confusion} (HMC)~\citep{maouche2018hmc} aims at preventing re-identification attacks by altering all the traces altogether.
The raw traces are transformed into heat maps, which are altered to look like another heat map in the dataset, and then transformed back to a \gls{gps} trace.

\paragraph{\textsc{Promesse}~\citep{promesse}} smooths the mobility traces, both temporally and geographically, to erase \glspl{poi} from the trace.
{\sc Promesse} ensures that, between each location sample, there is at least a given time and distance interval.
In the resulting mobility trace, the user appears to have a constant speed.
While {\sc Promesse} blurs the time notion from the trace---\emph{i.e.}, the user never appears to stay at the same place---it does not alter their spatial characteristics.
Yet, while \glspl{poi} may be still inferred if the user repeatedly goes to the same places, it will be harder to distinguish such \glspl{poi} from more random crossing points.

It is also possible to combine several \glspl{lppm} to improve the privacy of users~\citep{meftah2019fougere,eden}.
Because of potential remote leaks, the user should anonymize their trace locally before sharing it, which is how {\sc Eden}~\citep{eden} operates.
However, {\sc Eden} has not been deployed: it has only been simulated on ADB.
Even more so: despite their validity and to the best of our knowledge, no \gls{lppm} has been implemented in mobile devices.
This is partly due to the tight constraints of mobile devices, memory-wise notably: HMC~\citep{maouche2018hmc}, for instance, requires locally loading a large set of \gls{gps} traces to operate.

  \section{The \INTACT framework}\label{sec:approach}

In order to present the \acrfull{intact} framework, our two proposals will be presented in order: first, our time series modelisation tool \acrfull{flair}, and then, the \gls{poi} attack algorithm \acrfull{ds}.

\newcommand{\examplefigurewidth}{.3\linewidth}
\newcommand{\examplefigurespacing}{\hspace{10pt}}

\subsection{\FLI: Online Time Series Modeling}\label{sub:fli}

To overcome the memory constraints of mobile devices, we claim that efficient temporal storage solutions must be ported onto ubiquitous environments.
In particular, we advocate the use of data modeling, such as \acrfull{pla}~\citep{swab,grutzmacher2018time} or {\sc Greycat}~\citep{greycat}, to increase the storage capacity of mobile devices.
Therefore, we introduce \FLI, an online time series modeling algorithm based on an iterative and continuous \acrshort{pla} to store approximate models of data streams on memory-constrained devices, instead of storing all the raw data samples as state-of-the-art temporal databases do.

The intuition behind \FLI is that time series generally do not vary abruptly but rather follow linear tendencies, whether their value increases, decreases or remains the same over time.
Linear segments being suitable to represent linear tendencies, one can imagine representing a given time series as interconnected segments, of which each segment models as much original data points as possible, while keeping an error below a predefined threshold.
\FLI does just that: given a configuration parameter $\epsilon \in \mathbb{R}^{+*}$, it models any univariate time series as a series of segments (or interpolations), dropping original data points (or samples) as long as the following invariant is preserved:

\begin{quotation}\itshape
  All samples modeled by an interpolation maintain an error below the maximum threshold $\epsilon$.
\end{quotation}

\FLI is an online algorithm: each of the original time series' data points are inserted in order.
During this process, \FLI decides whether the latest sample can be modeled with the latest interpolation, or if a new segment needs to be created.
We will first present \FLI's representation of the time series, then how reads are performed, before diving into the insertion algorithm.

\paragraph{\FLI's mathematical formulation}
Consider an ever-growing univariate real-valued time series $\serie \subset \mathbb{R}^2$, such that its \textit{i}-th point is expressed as: $\serie\left[i\right] = \left( \abcisse_i, \ordonnee_i \right)$,
with $i \in \mathbb{N}^*$,
where $\abcisse_i \in \mathbb{R}^+$ is the point's timestamp, 
and $\ordonnee_i \in \mathbb{R}$ is its value.
Let \datastructure be the data structure containing \FLI's representation of \serie under an error bound of $\epsilon$. 
We say that \datastructure \emph{models} \serie under $\epsilon$, or: $\datastructure \models_\epsilon \serie$.

\begin{figure}[h]
  \center
  \includegraphics[width=.7\linewidth]{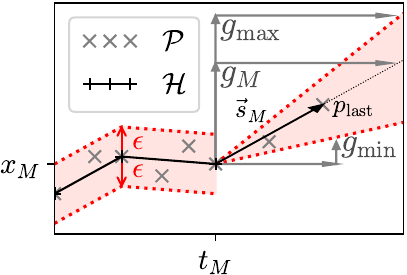}
  \caption{%
    A time series \serie being modeled with \FLI, displaying every component of \FLI's data structure $\datastructure = \left(\history, \acurr, \amin, \amax, \plast\right)$.
    \history models every point in \serie while keeping their error below $\epsilon$. 
    \history's last entry, $\pcurr = (\tcurr, \vcurr)$, is the origin of the latest model \scurr, with slope  \acurr.
    \scurr can notably be used to read past \pcurr, into the future.
    Gradients \amin and \amax bound the next sample insertions so as to keep their modeling error below $\epsilon$. 
    The three slopes are updated upon insert.
    The last inserted point \plast is also saved for next iterations.
  }
  \label{fig:approach_example4}
\end{figure}

The data structure $\datastructure$ is composed of {\em i)} a series of historical points $\history$ selected from \serie, {\em ii)} a triplet of gradients $\left(\acurr, \amin, \amax\right)$ that are used for reading and insertion, and {\em iii)} the last inserted point \plast.
How these elements are used is displayed in Fig.~\ref{fig:approach_example4}.
\begin{align*}
  &\datastructure = \left(\history, \acurr, \amin, \amax, \plast\right) \; \text{s. t.}\\
  &\left\{
  \begin{aligned}
    &\history \subset \serie \subset \mathbb{R}^2 \\
    &\left(\acurr, \amin, \amax\right) \in \mathbb{R}^3\\
    &\plast \in \mathbb{R}^2
  \end{aligned}
  \right.
\end{align*}

Any \emph{segment} $\segment$ in \history \emph{models} one or more points in the original time series \serie: 

\begin{gather}
  \left\{
  \begin{aligned}
    &\forall k \in \left[\!\left[1, \left|\history\right|\right[\!\right[\\
    &\exists \left(i, j\right) \in \left[\!\left[1, \left|\serie\right|\right]\!\right]^2\\
    &i < j
  \end{aligned}
  \right. \; \text{s. t.}
  \left\{
  \begin{aligned}
    &\history\left[k\right] = \serie\left[i\right] \\
    &\history\left[k+1\right] = \serie\left[j\right]\\
    &\segment_k = \left\langle \history\left[k\right], \history\left[k+1\right] \right\rangle
  \end{aligned}
  \right. \; \text{then}\nonumber\\[1ex]
  \forall l \in \left[\!\left[ i, j\right[\!\right[,
  \segment_k \models_\epsilon \serie\left[l\right] \iff
  \text{dist}\left(\segment_k, \serie\left[l\right]\right) < \epsilon
\end{gather}

The latest (or \emph{current}) segment $\scurr$ is treated differently: as it starts from the last point in \history, \FLI stores its slope $\acurr$ in \datastructure:

\begin{gather}
  \left\{
  \begin{aligned}
    &M = \left|\history\right|\\
    &\exists i \in \left[\!\left[1, \left|\serie\right|\right]\!\right]
  \end{aligned}
  \right. \; \text{s. t.}
  \left\{
  \begin{aligned}
    &\history\left[M\right] = \serie\left[i\right] \\
    &\scurr = \left\langle\history\left[M\right], \acurr\right\rangle\\
  \end{aligned}
  \right. \; \text{then}\nonumber\\[1ex]
  \forall l \in \left[\!\left[ i, \left|\serie\right|\right]\!\right],
  \scurr \models_\epsilon \serie\left[l\right] \iff
  \text{dist}\left(\scurr, \serie\left[l\right]\right) < \epsilon
\end{gather}

\paragraph{Reading data streams}
In \FLI, reading a value at time $\abcisse$ is achieved by estimating its image using the appropriate interpolation, as detailed in Algorithm~\ref{algo:read}.

If $\abcisse$ is lower than the last timestamp stored in the history ($\tcurr$ for short), then a position on an interpolation stored in \history will be retrieved.
Line~\ref{alg:read:search} selects the approriate index $k$ such that $\segment_k = \left\langle\history\left[k\right], \history\left[k+1\right]\right\rangle$ contains the queried timestamp: 
$\history\left[k\right].\abcisse \leq \abcisse < \history\left[k+1\right]$.
In practice, the lookup is done with a dichotomy search.
Lines~\ref{alg:read:slope}-\ref{alg:read:return} return the image of $\abcisse$ on the segment $\segment_k$, after having computed the gradient \gradient between the consecutive historical points $\history\left[k\right]$ and $\history\left[k+1\right]$.

If $\abcisse$ is ulterior or equal to $\tcurr$, line~\ref{alg:read:current2} returns the image on the current interpolation $\scurr = \left\langle\pcurr, \acurr\right\rangle$.

\begin{algorithm}[t]
\caption{Approximate read using \FLI}
\label{algo:read}
\begin{algorithmic}[1]
  \Require 
  $\left\{
  \begin{aligned}
    &\datastructure = \left(\history, \acurr, \amin, \amax, \plast\right)\\ 
    &M = \left|\history\right|\\
    &\left(\tcurr, \vcurr\right) = \pcurr
  \end{aligned}
  \right.$

  \Function{Read}{$\tqueried \in \mathbb{R}$}

  \If{$\abcisse \leq \tcurr$}
    \label{alg:read:historical}
    \Comment{Historical read}
    \State {\bf find}~$k \in \left[\!\left[1, M\right[\!\right[$ {\bf s. t.} 
    $\left\{
    \begin{aligned}
      &\history\left[k\right].\abcisse \geq \abcisse \\
      &\history\left[k+1\right].\abcisse < \abcisse \\
    \end{aligned}
    \right.$
    \label{alg:read:search}
    \State $\gradient \gets \frac{\history\left[k+1\right].\ordonnee - \history\left[k\right].\ordonnee}{\history\left[k+1\right].\abcisse - \history\left[k\right].\abcisse}$
    \Comment{Compute img. on $\segment_k$}
    \label{alg:read:slope}
    
    \State \Return $\gradient \times (\tqueried-\history\left[k\right].t) + \history\left[k\right].x$\label{alg:read:return}
  \Else
    \label{alg:read:current1}
    \Comment{Forward read}
    \State \Return $\acurr \times (\tqueried-\tcurr) + \vcurr$\label{alg:read:current2}
  \EndIf
  \EndFunction
\end{algorithmic}
\end{algorithm}

\paragraph{Inserting data samples}
Let us now go through the most important algorithm in \FLI, the sequential insertion, which is in charge of maintaining the invariant: \emph{all modeled points in \serie have an approximation error below $\epsilon$.}
Data samples are inserted sequentially: the current interpolation is adjusted to fit new samples until it cannot satisfy the invariant.
A naive solution to maintain the invariant while updating the current model would be to memorize every sample between $\pcurr$ and the last observed sample,
to check their error against the model---which would be costly.
Instead, \FLI only maintains $\amin$ and $\amax$, which are cost-effectively updated at each sample insertion.

Notation-wise, we will consider than the original series \serie is ever-growing, and that at each insertion, \FLI handles sample $\serie\left[\left|\serie\right|\right]$, noted $\pinserted = \left(\tinserted, \vinserted\right)$.
The last point inserted in \FLI's datastructure \datastructure is noted $\plast = \left(\tlast, \vlast\right)$.
Before insertion, \pinserted and \plast coexist.
After insertion, \plast takes the value of \pinserted.

\begin{figure}[t]
  \center
  \captionsetup[subfloat]{farskip=0pt}
  \subfloat[\pinserted's gradient $\ainserted$ remains within ${\lbrack}\amin,\amax{\rbrack}$.]{
    \includegraphics[width=.465\linewidth]{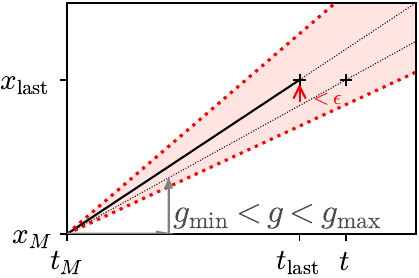}
    \label{fig:approach_example3A}
  }\hfill%
  \subfloat[The model is updated to encompass \pinserted.]{
    \includegraphics[width=.465\linewidth]{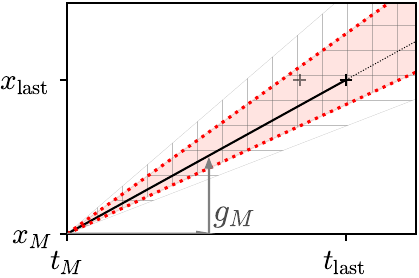}
    \label{fig:approach_example3B}
  }%
  \caption{%
    When a new sample fits within ${\lbrack}\amin,\amax{\rbrack}$, it is added to the current model by updating $\amin$ and $\amax$.
    $\acurr$ is also updated for read queries (see Alg.~\ref{algo:read}).
  }%
  \label{fig:approach_example3}
\end{figure}

\begin{figure}[t]
  \center
  \captionsetup[subfloat]{farskip=0pt}
  \subfloat[The error of \pinserted exceeds $\epsilon$.]{
    \label{fig:approach_example2A}
    \includegraphics[width=.465\linewidth]{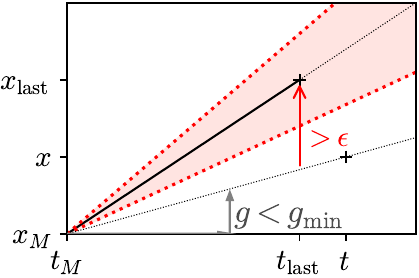}
  }\hfill%
  \subfloat[A new model is created from the last two points.]{
    \label{fig:approach_example2B}
    \includegraphics[width=.465\linewidth]{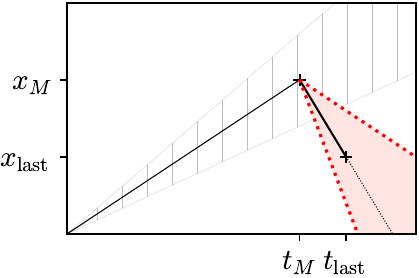}
  }%
  \caption{
    When an error $> \epsilon$ is reported, a new model is created using $\plast$ as its origin.
  }%
  \label{fig:approach_example2}
\end{figure}

Following along Algorithm~\ref{algo:insert}, we will first cover the insertion of samples that fall within the current interpolation (represented in Fig.~\ref{fig:approach_example3}), before explaining how a new segment is generated once the last point breaks the invariant (displayed in Fig.~\ref{fig:approach_example2}).

Upon insertion of a new sample \pinserted, lines~\ref{alg:ins:at1}-\ref{alg:ins:at2} first compute the slope \ainserted of the segment $\segment = \left\langle\pcurr, \pinserted\right\rangle$.
On line~\ref{alg:ins:if}, \ainserted is compared to the interval $\left]\amin, \amax\right[$, to know whether \pinserted falls within the bounds of the current interpolation \scurr or not.

\begin{algorithm}[t]
\caption{Insertion using $\epsilon \in \mathbb{R}^{+*}$}
\label{algo:insert}
\begin{algorithmic}[1]
  \Require 
  $\left\{
  \begin{aligned}
    &\datastructure = \left(\history, \acurr, \amin, \amax, \plast\right)\\ 
    &M = \left|\history\right|\\
    &\left(\tcurr, \vcurr\right) = \pcurr
  \end{aligned}
  \right.$

  \Function{insert}{$\pinserted = \left(\tinserted, \vinserted\right) \in \mathbb{R}^2, \; \epsilon$}

    \State $\left(\tnorm, \vnorm\right) \leftarrow  \left(\tinserted - \tcurr, \vinserted - \vcurr\right)$\label{alg:ins:at1}
    \State $ \ainserted \leftarrow  \vnorm / \tnorm$\label{alg:ins:at2}
    \Comment{Compute $\ainserted$}

    \If{$\amin < \ainserted < \amax$}
      \label{alg:ins:if}

      \State $\acurr \leftarrow \ainserted$\label{alg:ins:update1}
      \Comment{Update model (Fig.~\ref{fig:approach_example3})}
      \State $\left(\amin^\pinserted, \amax^\pinserted\right) \leftarrow \left( \frac{\vnorm - \epsilon}{\tnorm}, \frac{\vnorm + \epsilon}{\tnorm} \right)$
      \label{alg:ins:cone}
      \State $\amin \leftarrow  \max\left(\amin,\amin^\pinserted\right)$
      \label{alg:ins:amin}
      \State $\amax \leftarrow  \min\left(\amax,\amax^\pinserted\right)$
      \label{alg:ins:amax}

    \Else

      \State $\history \gets \history \cup \left[\left(\tlast, \vlast\right)\right]$ %
      \Comment{New model (Fig.~\ref{fig:approach_example2})}\label{alg:ins:persist}
    
      \State $\left(\tnorm', \vnorm'\right) \leftarrow  \left(\tinserted - \tlast, \vinserted - \vlast\right)$ \label{alg:ins:new1}
      \State $\acurr \leftarrow  \vnorm' / \tnorm'$ 
      \State $\amin \leftarrow  \left(\vnorm'-\epsilon\right)/\tnorm'$
      \State $\amax \leftarrow  \left(\vnorm'+\epsilon\right)/\tnorm'$\label{alg:ins:new2}

    \EndIf

    \State $\left(\tlast, \vlast\right) \leftarrow  \left(\tinserted, \vinserted\right)$
    \Comment{Update~last~sample}\label{alg:ins:penultimate}
  \EndFunction
\end{algorithmic}
\end{algorithm}

If it falls within (cf. Fig.~\ref{fig:approach_example3}), \pinserted is added to the current interpolation: 
\acurr is updated to \ainserted on line~\ref{alg:ins:update1}, \acurr being used for reading ahead (see Alg.~\ref{algo:read}). 
Two new slopes $\amin^\pinserted$ and $\amax^\pinserted$ are computed on line~\ref{alg:ins:cone}. 
They respectively represent the lines $\left\langle\pcurr, \left(\tinserted, \vinserted - \epsilon\right)\right\rangle$ and $\left\langle\pcurr, \left(\tinserted, \vinserted + \epsilon\right)\right\rangle$.
Together, those two lines materialise \pinserted's \emph{allowed cone} for future values, so as to keep \pcurr as the origin of the model while preserving the invariant for \pinserted and the future insert. 
On line~\ref{alg:ins:amin}, \amin takes the maximum between its previous value and $\amin^\pinserted$; 
on line~\ref{alg:ins:amax}, the converse operation happens to \amax.
What these two lines do is \emph{shrinking} the interval $\left] \amin, \amax \right[$ so as to encompass \pinserted's allowed cone in the next iteration.
Recursively, \amin and \amax represent the current model's allowed cone, which is \emph{the intersection of all modeled points' allowed cones}---similarly to \cite{galakatos2019fiting}, but in an online fashion.
The process is pictured in Fig.~\ref{fig:approach_example3B}: vertical hatches represent the previous point's allowed cone, horizontal ones show the last point's, while the model's allowed cone is represented in red.

If \ainserted falls outside $\left] \amin, \amax \right[\;$(that is: if \pinserted falls outside the model's allowed cone), then a new interpolation must begin from the two last points (see Fig.~\ref{fig:approach_example2}).
It begins from the last two observed samples: the new $\scurr$ becomes $\left\langle\plast, \pinserted\right\rangle$.
The penultimate sample $\plast$ is thus persisted to the series of selected points $\history$ at line~\ref{alg:ins:persist}, and new values for $\acurr, \amin$ and $\amax$ are derived based on the new $\scurr$ (lines~\ref{alg:ins:new1}--\ref{alg:ins:new2}).
Recognize here how the computation of the new gradients is equivalent to the computation of \pinserted's allowed cone on line~\ref{alg:ins:cone} (with a different origin).

In any case, the penultimate sample $\plast$ is updated on line~\ref{alg:ins:penultimate}.


\paragraph{The case of $\epsilon$}
The value of the error bound parameter has an important impact on the performances of \FLI.
If $\epsilon$ is too small, none of the inserted samples fit the current model at that time, thus initiating a new model each time.
In that case, there will be one model per sample, imposing an important memory overhead.
The resulting model overfits the data.
On the other hand, if $\epsilon$ is too large, then all the inserted samples fit, and a single model is kept.
While it is the best case memory-wise, the resulting model simply connects the first and last points and underfits the data.

The choice of $\epsilon$ is highly dependent on the input data. 
For instance, accelerometer data---that generally has a high sampling rate and little variance---are modeled efficiently even with small values of $\epsilon$.
Whereas GPS data---slow sampling rate and high variance---require finer tuning.
We propose a parameter tuning process in Section~\ref{ssec:parameter_tuning}.

On the other hand, the cautious reader will have observed how $\epsilon$ is only a parameter of the \textsc{Insert} function in Alg.~\ref{algo:insert}. 
In future works, our team intends to dynamically adapt this error bound depending on the data and its freshness.

\subsection{\STAY: Attacking Location Privacy}
\label{sub:stay}
\newcommand{\minTime}{t_{min}}
\newcommand{\maxDistance}{d_{max}}
\newcommand{\maxSize}{s_{max}}
\newcommand{\Stays}{S}
\newcommand{\geodist}{\text{dist}_\mathbb{G}}

Time series modeling being out of the way, we will now present our mobile-ready geolocation privacy attack tool: \acrfull{ds}.
The purpose of such a tool is to audit the privacy of mobility traces directly where they are generated: on resource-constrained devices such as smartphones. 
Only then would it be reasonable privacy-wise to upload one's location traces to a third-party.
\STAY constitutes an enhancement of the original \poiattack algorithm proposed by \cite{primault2014differentially}.
The original contribution being too costly to run on mobile, \STAY proposes a mobile-ready variation.
Just like \FLI, \STAY fits into the broader \INTACT framework, the mobile privacy system that constitutes our whole contribution.

In \poiattack, the \gls{poi} disclosure is done by a two-steps algorithm:
potential candidates for \glspl{poi} (dubbed \emph{stays}) are first extracted,
then \emph{stays} are merged to avoid duplication of similar \glspl{poi}.
A \emph{stay} is defined as a circle with a radius lower than $\maxDistance$ where a user spent a time higher than a set time $\minTime$.
A \emph{stay} is represented by its center. %
The two thresholds $\minTime$ and $\maxDistance$ have an important impact on the type of \acrshort{poi} extracted.
\emph{Short stays} will identify day-to-day patterns, such as shopping preferences, while \emph{long stays} will identify \emph{e.g.} travel preferences.
In the second step, the \emph{stays} whose centroids are close enough are merged to obtain the final list of \acrshortpl{poi}.
\poiattack~\citep{primault2014differentially} iterates linearly over the mobility trace and compute \emph{stays} as they appear.
This approach is expensive for denser mobility traces---\emph{i.e.}, with high frequency sampling.
It is prohibitively long to execute on constrained devices like mobile phones.

Our contribution \stay (\STAY) instead proposes a divide-and-conquer strategy that scales with the data density.
The intuition behind \STAY is to avoid wasting time looking for \emph{stays} in portions of the trace where they are impossible,
\emph{i.e.} where more than $\maxDistance$ has been traveled in less than $\minTime$,
\emph{e.g.} a car trip at high speed in a straight line.
While the regular approach would consider each location until the end of the trace, \STAY skips it entirely.
The key idea of \stay is to recursively divide the trace until either such a \emph{stay}-less segment is found and discarded, or until a fixed size segment is found on which the regular way to extract \emph{stays} is performed.

Algorithm~\ref{algo:stays} depicts the pseudo-code of \STAY.
It scrutinizes the GPS trace $T \in \left(\mathbb{R} \times \mathbb{G}\right)^n$, composed of $n$ samples.
For each $i \in \llbracket 0, n-1 \rrbracket$, $T\left[i\right].t$ represents the $i_\text{th}$ sample's timestamp, while $T\left[i\right].g$ is its position in whatever geographical space $\mathbb{G}$ equipped with a distance function $\geodist$.
\STAY has three configuration parameters: the aforementioned $\minTime$ and $\maxDistance$ representing the time and space limits of a \gls{poi}, and $\maxSize$, the sub-trace size threshold below which the divide-and-conquer approach for POI inference is abandoned in favor of the iterative one.
Three indices are manipulated, all called $i$ with a self-explanatory subscript.
The \STAY function takes $i_\text{first}$ and $i_\text{last}$ as arguments, being the bounds of the sub-trace under study.
On the first call, the whole input space is provided: $i_\text{first}$ is 0 and $i_\text{last}$ takes $n-1$.
Subsequent recursive calls provide either the first half of the input sub-trace, or the second, until an iterative search is preferred.

On lines~\ref{alg:stay:iter1}~to~\ref{alg:stay:iter2}, the size of the input sub-trace is checked against the threshold $\maxSize$.
If the trace is smaller, then a linear search for \emph{stays} is performed \emph{à la} \poiattack~\citep{primault2014differentially}.
On l.~\ref{alg:stay:split}, the indices space is split: $i_\text{split}$ is set to the midpoint between $i_\text{first}$ and $i_\text{last}$.
Lines~\ref{alg:stay:left1}-\ref{alg:stay:left2} check whether the left sub-trace $T\left[i_\text{first}..i_\text{split}\right]$ is susceptible to contain \emph{stays}, in which case \STAY is recursively called.
Its output fills the list of stays $\Stays$.
As already mentioned, a sub-trace cannot contain any \emph{stay} if a distance of more than $\maxDistance$ was traveled in less than $\minTime$.
Lines~\ref{alg:stay:right1}~to~\ref{alg:stay:right2} perform the same check for the right sub-trace $T\left[i_\text{split}..i_\text{last}\right]$, in which case the result of the recursive call is added to $\Stays$.
Finally, $\Stays$ is returned. %
\poiattack's merge of \emph{stays} into \glspl{poi} must be subsequently performed.

The more discarded segments, the faster compared to the regular approach.
\emph{Stays} around the midpoints $i_\text{split}$ could be missed, but \STAY ignores them because a \acrshort{poi} is a cluster of several stays:
it is very unlikely to miss them all.
\STAY can be implemented sequentially or concurrently, to leverage multi-core processors.

\begin{algorithm}[t]
\caption{\stay (\STAY) using parameters $(\minTime, \maxDistance, \maxSize) \in \mathbb{R}^{3+}$}
\label{algo:stays}
\begin{algorithmic}[1]
  \Require $T\in{(\mathbb{R} \times \mathbb{G})^n}$

  \Function{\STAY}{$\left(i_\text{first}, i_\text{last}\right) \in \llbracket 0, n-1 \rrbracket^2$}

    \If{$i_\text{last} - i_\text{first} \le \maxSize$} \Comment{Iterative case}
    \label{alg:stay:iter1}
    \State \Return $\text{getStays}\left(T\left[i_\text{first}..i_\text{last}\right]\right)$
    \EndIf
    \label{alg:stay:iter2}

    \State $\Stays \leftarrow \emptyset$
    \State $i_\text{split} \gets \lfloor \left(i_\text{first} + i_\text{last}\right) / 2 \rfloor$
    \label{alg:stay:split}

    \Statex \Comment{Left sub-trace recursion}
    \State $t_\Delta \gets T\left[i_\text{split}\right].t-T\left[i_\text{first}\right].t$ 
    \label{alg:stay:left1}
    \State $d_\Delta \gets \geodist\left(T\left[i_\text{first}\right].g, T\left[i_\text{split}\right].g\right)$
    \If{$\lnot \left(d_\Delta > \maxDistance \land t_\Delta \leq \minTime \right)$}
      \State $\Stays \gets \text{\STAY}\left(i_\text{first}, i_\text{split}\right)$
    \EndIf
    \label{alg:stay:left2}

    \Statex \Comment{Right sub-trace recursion}
    \label{alg:stay:right1}
    \State $t_\Delta \gets T\left[i_\text{last}\right].t-T\left[i_\text{split}\right].t$ 
    \State $d_\Delta \gets \geodist\left(T\left[i_\text{split}\right].g,T\left[i_\text{last}\right].g\right)$
    \If{$ \lnot \left(d_\Delta > \maxDistance \land t_\Delta \leq \minTime \right)$}
      \State $\Stays \gets \Stays \cup \text{\STAY}\left(i_\text{split}, i_\text{last}\right)$
    \EndIf
    \label{alg:stay:right2}

    \State \Return $\Stays$
    \label{alg:stay:ret}
  \EndFunction
\end{algorithmic}
\end{algorithm}

\section{Experimental Setup}\label{sec:setup}
\newcommand{\phone}{Fairphone 3\xspace}
\newcommand{\phoneDetails}{Qualcomm Snapdragon\,632 with 4GB of RAM}
\newcommand{\phoneAndroid}{Android\,11\xspace}
\newcommand{\iphone}{iPhone 12\xspace}
\newcommand{\iphoneDetails}{Apple~A14 Bionic~SoC with with 4GB of RAM}
\newcommand{\iphoneVersion}{iOS\,15.1.1\xspace}

\subsection{Key Performance Metrics}

\emph{Memory footprint.}
The key objective of \FLI is to reduce the memory footprint required to store an unbounded stream of samples.
We explore two metrics: {\it(i)} the number of 64-bit variables required by the model and {\it(ii)} the size of the model in the device memory.
To do so, we compare the size of the persistent file with the size of the vanilla {\sc SQLite} database file.
We consider the number of 64-bit variables as a device-agnostic estimation of the model footprint.

\emph{I/O throughput.}
Another key system metric is the I/O throughput of the temporal databases.
In particular, we measure how many write and read operations can be performed per second (IOPS).

\emph{\gls{poi} quality.} %
Measuring the quality of inferred \glspl{poi} is difficult, as there is no acknowledged definition of how to compute \glspl{poi}.
We consider as our ground truth the \glspl{poi} inferred by the state-of-the-art \gls{poi}-attack~\citep{primault2014differentially}, which we refer to as the `raw' \glspl{poi}.
The existence of such a `ground-truth' is however debatable, as two different---but close---\glspl{poi} can be merged by the algorithm into a single \gls{poi}.
As an example, if a user visits two different shops separated by a road, but their distance is lower than $d_{max}$, those will be merged into a single \gls{poi} located at the center of the road.
For that reason, we need two metrics to compare the sets of \glspl{poi} returned in the different cases: the distance between \glspl{poi}, and the sets' sizes.

\paragraph{Distance between \glspl{poi}.}
As the \gls{poi} definition is mainly algorithmic, we compute the distance of each obtained \gls{poi} to its closest raw \gls{poi} as the metrics assessing the quality of new \glspl{poi}.
These distances are reported as a \gls{cdf}.
If \FLI does not alter significantly the locations of the mobility traces it captures, the computed distances should be short.

\paragraph{Number of \glspl{poi}.}
In addition to the distances between \glspl{poi}, we are also considering their returned quantity as a metric.
In our previous example, visiting the two shops may result in two different \glspl{poi} because they have been slightly shifted by \FLI.
Beyond the numbers, we expect that {\sc Promesse} successfully anonymizes mobility traces by returning a total of zero \gls{poi}.

\subsection{Input Datasets \& Parameter Tuning}
\label{ssec:parameter_tuning}

\begin{figure*}
  \captionsetup[subfloat]{farskip=0pt}
  \begin{center}
    \subfloat[\acrshort{cdf} of {\sc Cabspotting} locations' drifts.]{
      \includegraphics[width=.3\linewidth]{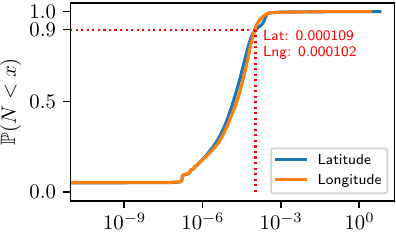}
      \label{fig:cabspotting_deltas}
    }%
    \qquad
    \subfloat[\acrshort{cdf} of {\sc PrivaMov} locations' drifts.]{
      \includegraphics[width=.3\linewidth]{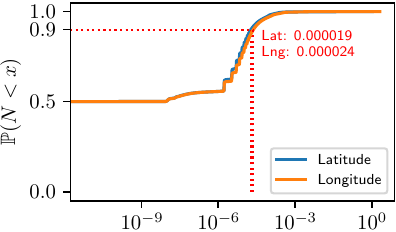}
      \label{fig:privamov_deltas}
    }
  \end{center}
  \caption{\acrfull{cdf} of latitude and longitude drifts of successive location samples in {\sc Cabspotting} and {\sc PrivaMov} datasets.
  One can observe that, from one location sample to the next, latitude or longitude deviations are small.}
  \label{fig:drift}
\end{figure*}

\FLI was built to allow the storage of user-generated data series, such as GPS and accelerometer streams (although it is readily applicable to other types of real-valued streams, as shown in Section~\ref{ssec:beyond_location_streams}).
However, the choice of the $\epsilon$ parameter depends on the underlying data distribution. 
Towards that end, we propose a semi-automated $\epsilon$ parameter selection routine.

In this section, we present the two datasets that will be used throughout our evaluation, before outlining and applying our parameter selection algorithm to the both of them.

\paragraph*{Location datasets.}
We use two real-world mobility datasets that display different characteristics:
\begin{itemize}
  \item \textsc{Cabspotting}~\citep{piorkowski2009crawdad} is a mobility dataset of $536$ taxis in the San Francisco Bay Area.
    The data was collected during a month and is composed of $11$ million records, for a total of 388\,MB.
    It is composed only of car trips in a dense urban environment.
  \item {\sc PrivaMov}~\citep{privamov} is a multi-sensor mobility dataset gathered during $15$ months by $100$ users around the city of Lyon, France. 
    It contains several transportation modes, most notably pedestrian.
    {\sc PrivaMov} displays a much higher sampling rate that {\sc Cabspotting}.
    We use the full \gls{gps} dataset, which includes $156$ million records, totaling 7.2\,GB.
\end{itemize}

\paragraph*{Parameter tuning.}
The choice of an $\epsilon$ value is of major importance and plays a central role in {\FLI}'s performance: 
a poorly-chosen value has a strong impact on {\FLI}'s underlying segments, either degrading modeled data quality or filling storage space up excessively.
To find a compromise between the two, since the $\epsilon$ value is highly correlated to the modeled data, one has to \textit{know} the data; more specifically, we propose to study the signal's amplitude variation between consecutive values.

For example, in the context of location data, Fig.~\ref{fig:drift} characterizes---as a \gls{cdf}---the evolution of longitude and latitude samples for all the traces stored in the {\sc Cabspotting} and {\sc PrivaMov} datasets.
In particular, we plot the \gls{cdf} of the drift $d$ observed between 2 consecutive values $(\abcisse_1,\ordonnee_1)$ and $(\abcisse_2,\ordonnee_2)$, which we compute as $d=|\ordonnee_2-\ordonnee_1|/|\abcisse_2-\abcisse_1|$.
One can observe that {\sc Cabspotting} and {\sc PrivaMov} datasets report on a drift lower than $1\times10^{-4}$ and $2\times10^{-5}$ for 90\% of the values, respectively.
Furthermore, due to the high density of locations captured by {\sc PrivaMov}, half of the drifts are equal to 0, meaning several consecutive longitudes or latitudes are unchanged.
This preliminary analysis highlights that \FLI can indeed efficiently model mobility data, and demonstrates that $\epsilon=10^{-3}$ is a conservative choice to model these datasets.

To automate the tuning of $\epsilon$, one must input a sample of their data stream into our provided script\footnote{See file \texttt{lib/epsilon\_choice.dart} in our code repository~\citep{flairimplementation}.}, which proposes candidate $\epsilon$ parameter values to capture 90\%, 95\% and 99\% of the sampled data.
We discuss perspectives for parameter selection in Section~\ref{sec:discussion}.

\subsection{Storage Competitors}\label{sec:competitors}
\textsc{SQLite} is the state-of-the-art solution to persist and query large volumes of data on Android devices.
{\sc SQLite} provides a lightweight relational database management system.
{\sc SQLite} is not a temporal database, but is a convenient and standard way to store samples persistently on a mobile device.
Insertions are atomic, so one may batch them to avoid one memory access per insertion.

{\em Sliding-Window And Bottom-up} (SWAB)~\citep{swab} is a linear interpolation model.
As \FLI, the samples are represented by a list of linear models.
In particular, reading a sample is achieved by iteratively going through the list of models until the corresponding one is found and then used to estimate the requested value.
The bottom-up approach of SWAB starts by connecting every pair of consecutive samples and then iterates by merging the less significant pair of contiguous interpolations.
This process is repeated until no more pairs can be merged without introducing an error higher than $\epsilon$.
Contrarily to \FLI, this bottom-up approach is an offline one, requiring all the samples to be known.
SWAB extends the bottom-up approach by buffering samples in a sliding window.
New samples are inserted in the sliding window and then modeled using a bottom-up approach: whenever the window is full, the oldest model is kept and the captured samples are removed from the buffer.

One could expect that the bottom-up approach delivers more accurate models than the greedy \FLI, even resulting in a slight reduction in the number of models and faster readings.
On the other hand, sample insertion is more expensive than \FLI due to the execution of the bottom-up approach when storing samples.
Like \FLI, SWAB ensures that reading stored samples is at most $\epsilon$ away from the exact values.

\textsc{Greycat}~\citep{greycat} aims at compressing even further the data by not limiting itself to linear models.
{\sc Greycat} also models the samples as a list of models, but these models are polynomials.
The samples are read the same way.

When inserting a sample, it first checks if it fits the model.
If so, then nothing needs to be done.
Otherwise, unlike \FLI and SWAB which directly initiate a new model, {\sc Greycat} tries to increase the degree of the polynomial to make it fit the new sample.
To do so, {\sc Greycat} first regenerates $d+1$ samples in the interval covered by the current model, where $d$ is the degree of the current model.
Then, a polynomial regression of degree $d+1$ is computed on those points along the new one.
If the resulting regression reports an error lower than $\frac{\epsilon}{2^{d+1}}$, then the model is kept, otherwise, the process is repeated by incrementing the degree until either a fitting model is found or a maximum degree is reached.
If the maximum degree is reached, the former model is stored and a new model is initiated.
The resulting model is quite compact, and thus faster to read, but at the expense of an important insertion cost.

Unlike \FLI and SWAB, there can be errors higher than $\epsilon$ for the inserted samples, as the errors are not computed on raw samples but on generated ones, which may not coincide.
Furthermore, the use of higher-degree polynomials makes the implementation subject to overflow: to alleviate this effect, the inserted values are normalized.

\subsection{Experimental Settings}
For experiments with univariate data streams---\emph{i.e.} memory and throughput benchmarks---we set $\epsilon=10^{-2}$.
The random samples used in those experiments follow a uniform distribution in $[-1{,}000;1{,}000]$: it is very unlikely to have two successive samples with a difference lower than $\epsilon$, hence reflecting the worst case conditions for \FLI.
For experiments on location data, and unless said otherwise, we set $\epsilon=10^{-3}$ for \FLI, SWAB and {\sc Greycat}.
For {\sc Greycat}, the maximum degree for the polynomials is set to $14$.
The experiments evaluating the throughput were repeated 4 times each and the average is taken as the standard deviation was low.
All the other experiments are deterministic and performed once.

\subsection{Implementation Details}
We implemented \FLI using the Flutter \emph{Software Development Kit} (SDK)~\citep{flutterframework}.
Flutter is Google's UI toolkit, based on the Dart programming language, that can be used to develop natively compiled apps for Android, iOS, web and desktop platforms (as long as the project's dependencies implement cross-compilation to all considered platforms).
Our implementation includes \FLI and its storage competitors.%
This implementation is publicly available~\citep{flairimplementation}.

For our experiments, we also implemented several mobile applications based on this library.
To demonstrate its capability of operating across multiple environments (models, operating systems, processors, memory capacities, storage capacities), all our benchmark applications were successfully installed and executed in the devices listed in Table~\ref{tab:test_devices}.
Unless mentioned otherwise, the host device for the experiments is the Fairphone\,3.

\begin{table*}
    \centering
    \caption{
        \label{tab:test_devices}
        Mobile devices used in the experiments.
    }
\resizebox{0.7\linewidth}{!}{
    \begin{tabular}{>{\columncolor{Gray}}l c cc rr}
        \hline
        \bf Model        & \bf OS      & \bf CPU    & \bf Cores & \bf RAM & \bf Storage \\ \hline
        Lenovo\,Moto\,Z  & Android\,8  & Snapdragon\,820    & 4 & 4\,GB  & 32\,GB \\
        Fairphone\,3     & Android\,11 & Snapdragon\,632    & 8 & 4\,GB  & 64\,GB\\
        Pixel\,7\,Pro    & Android\,13 & Google\,Tensor\,G2 & 8 & 12\,GB & 128\,GB\\
        iPhone\,12       & iOS\,15.1.1 & A14\,Bionic        & 6 & 4\,GB  & 64\,GB\\
        iPhone\,14\,Plus & iOS\,16.0.1 & A15\,Bionic        & 6 & 6\,GB  & 128\,GB\\
        \hline
    \end{tabular}
}%
\end{table*}

  \section{Experimental Results}
\label{sec:eval}
In this section, we evaluate our implementation of \FLI{} on Android and iOS to show how it enables efficient data stream storage on mobile devices. %
We first perform several benchmarks (memory, throughput \& stability), before evaluating the performance of \FLI{} beyond location streams.
Finally, we perform a \gls{poi} mining experiment directly on mobile devices, thus showcasing how \INTACT enables \emph{in-situ} big data processing.

\subsection{Memory Benchmark}
As there is no temporal database (e.g. {\sc InfluxDB}), available on Android, we compare \FLI's performances with {\sc SQLite}, the only database natively available on Android.

\emph{Synthetic data}. 2 identical operations are performed with {\sc SQLite} and \FLI: {\it (i)} the incremental insertion of random samples and {\it (ii)} the incremental insertion of constant samples.
The memory footprint of both solutions on disk is compared when storing timestamped values.
As \FLI{} models the inserted samples, random values are the worst-case scenario it can face, while inserting constant values represents the ideal one.
One million samples are stored and, for every $10{,}000$ insertion, the size of the file associated with the storage solution is saved.
The experiments are done with a publicly available application~\citep{benchmarkingmemoryspace}.

\begin{figure}
    \centering
    \includegraphics[width=0.6\linewidth]{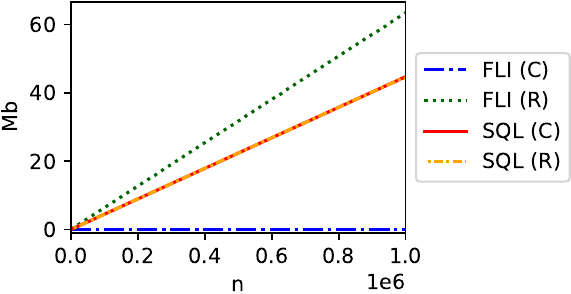}
    \caption{Inserting $1M$ samples, random ({\sf R}) or constant ({\sf C}), in {\sc SQLite} and \FLI.}
    \label{fig:xp_benchmarking}
    \centering
    \includegraphics[width=0.5\linewidth]{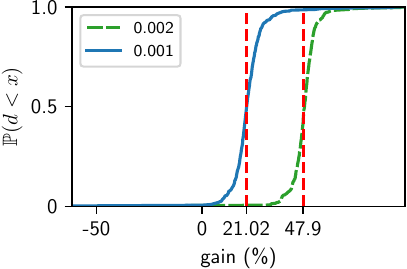}
    \caption{Memory gain distribution when storing {\sc Cabspotting} with \FLI.
    }
    \label{fig:xp_gains}
\end{figure}

Fig.~\ref{fig:xp_benchmarking} depicts the memory footprint of both approaches.
On the one hand, the size of the {\sc SQLite} file grows linearly with the number of inserted samples, no matter the nature (random or constant) of the samples.
On the other hand, the \FLI{} size grows linearly with random values, while the size is constant for constant values.
In particular, for the constant values, the required size is negligible.
The difference between vanilla {\sc SQLite} and \FLI{} is explained by the way the model is stored: while {\sc SQLite} optimizes the way the raw data is stored, \FLI{} is an in-memory stream storage solution, which naively stores coefficients in a text file.
Using more efficient storage would further shrink the difference between the two.
As expected, the memory footprint of a data stream storage solution outperforms the one of a vanilla {\sc SQLite} database in the case of stable values.
While random and constant values are extreme cases, in practice data streams produced by ubiquitous devices exhibit a behavior between the two scenarios which allows \FLI{} to lower the memory required to store those data streams.

\emph{GPS data}. We use \FLI{} to store latitudes and longitudes of the entire {\sc Cabspotting} dataset ($388$MB) in memory, using both $\epsilon = 10^{-3}$ and $\epsilon = 2 \times 10^{-3}$ (representing an accuracy of approximately a hundred meters).
For each user, we compute the gain of memory storage as a percentage, compared to storing the raw traces.
Fig.~\ref{fig:xp_gains} reports on the gain distribution as a \gls{cdf} along with the average gain on the entire dataset.
Most of the user traces largely benefit from using \FLI, and \FLI{} provides an overall gain of $21$\% ($307$MB) for $\epsilon = 10^{-3}$ on the entire dataset, and a gain of $47.9\%$ ($202$MB) for $\epsilon = 2 \times 10^{-3}$.

Additionally, we also compare {\sc SQLite} and \FLI{} to store the entire {\sc Privamov} dataset ($7.2$GB).
In this context, \FLI{} only requires $25$MB (gain of $99.65$\%) compared to more than $5$GB (gain of $30.56$\%) for {\sc SQLite}, despite the naive storage scheme used by \FLI.
Furthermore, with smartphones featuring limited RAM~(cf. Table~\ref{tab:test_devices}) and not allocating the whole of it to a single application, \FLI{} enables loading complete datasets in memory to be processed: on mobile devices, loading the raw {\sc Privamov} dataset in memory crashes the application (due to out-of-memory errors), while \FLI{} succeeds in fitting the full dataset into RAM.
This capability is particularly interesting to enable the deployment of data stream processing tasks on mobile devices that do not incur any processing overhead.

\subsection{Throughput Benchmark}
We compare \FLI{} with its competitors among the temporal databases: SWAB and {\sc Greycat}.
We study the throughput of each approach in terms of IOPS.
Insertion speed is computed by inserting $1M$ random samples (that is each of these solutions' worst-case scenario).
For the reads, we also incrementally insert $1M$ samples before querying $10K$ random samples among the inserted ones.
{\sc Greycat} is an exception: due to its long insertion time (Sect.~\ref{sec:competitors}), we only insert $10K$ random values and those values are then queried.
Our experiment is done using a publicly available application~\citep{benchmarkingthroughput}.

\begin{figure}
  \captionsetup[subfloat]{farskip=0pt}  \center
  \subfloat[Sequential Insertions]{
    \includegraphics[width=.475\linewidth]{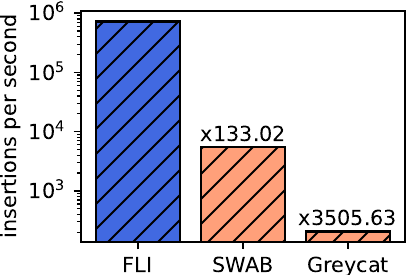}
    \label{fig:xp_benchmarkingthroughput_insertions}
  }
  \subfloat[Random Reads]{
    \includegraphics[width=.475\linewidth]{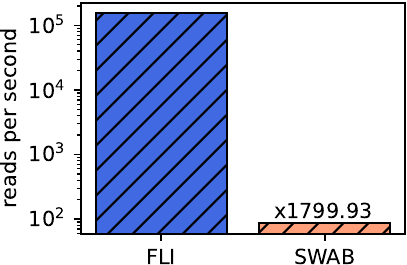}
    \label{fig:xp_benchmarkingthroughput_reads}
  }
  \caption{
    Throughput for insertions and reads using \FLI, SWAB, and {\sc Greycat} (log scale). %
    \FLI{} drastically outperforms its competitors for insertions and reads. %
  }
  \label{fig:xp_benchmarkingthroughput}
\end{figure}

Fig.~\ref{fig:xp_benchmarkingthroughput} depicts the throughput of the approaches for sequential insertions and random reads.
On the one hand, \FLI{} drastically outperforms its competitors for the insertions: it provides a speed-up from $\times\,133$ against SWAB up to $\times\,3{,}505$ against {\sc Greycat}.
The insertion scheme of \FLI{} is fast as it relies on a few parameters.
On the other hand, {\sc Greycat} relies on a costly procedure when a sample is inserted: it tries to increase the degree of the current model until it fits with the new point or until a maximum degree is reached.
{\sc Greycat} aims at computing a model as compact as possible, which is not the best choice for fast online insertions.

\newcommand{\scaleBenchmarking}{0.58}

For the reads (Fig.~\ref{fig:xp_benchmarkingthroughput_reads}), \FLI{} also outperforms SWAB. %
Our investigation reports that \FLI{} largely benefits from its dichotomy lookup inside the time index (see Alg.~\ref{algo:read}),
compared to SWAB, which scans the list of models sequentially until the correct time index is found.
SWAB reads have a complexity linear in the size of the list, while \FLI{} has a logarithmic one.
{\sc Greycat} has the same approach as SWAB and this is why it is not represented in the results: with only $10K$ insertions instead of $1M$, its list of models is significantly smaller compared to the others, making the comparison unfair.
Nevertheless, we expect {\sc Greycat} to have a better throughput as its model list shall be shorter.

Note that those results have been obtained with the worst-case: random samples.
Similarly, unfit for \FLI{} are periodical signals, such as raw audio: our tests show a memory usage similar to random noise.
Because \FLI{} leverages linear interpolations, it performs best with signals that have a linear shape (e.g. \gls{gps}, accelerometer).
We expect SWAB to store fewer models than \FLI{} thanks to its sliding window, resulting in faster reads.
However, the throughput obtained for \FLI{} is minimal and \FLI{} is an order of magnitude faster than SWAB for insertions, so it does not make a significant difference.
We can conclude that \FLI{} is the best solution for storing large streams of data samples on mobile devices.

\subsection{Stability Benchmark}
\begin{figure}
  \captionsetup[subfloat]{farskip=0pt}
  \center
  \subfloat[Samples per {\sc Cabspotting} model.]{
    \includegraphics[width=.475\linewidth]{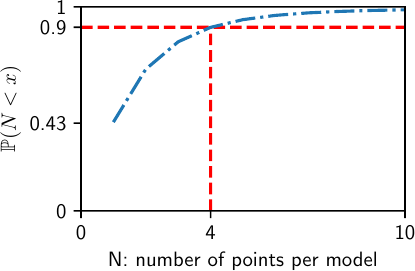}
    \label{fig:stability_cabspotting_samples}
    }
    \subfloat[Duration of {\sc Cabspotting} models.]{
      \includegraphics[width=.475\linewidth]{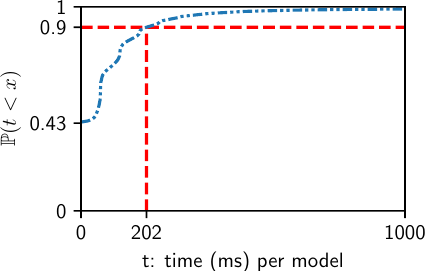}
      \label{fig:stability_cabspotting_time}
    }
  \vspace{1pt}
  \subfloat[Samples per {\sc PrivaMov} model.]{
    \includegraphics[width=.475\linewidth]{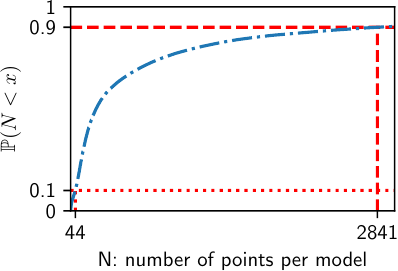}
    \label{fig:stability_privamov_samples}
    }
    \subfloat[Duration of {\sc PrivaMov} models.]{
      \includegraphics[width=.475\linewidth]{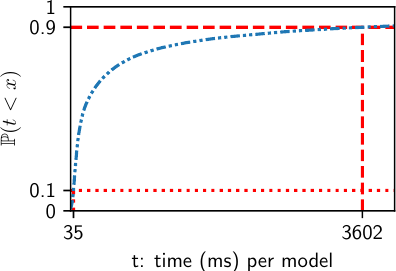}
      \label{fig:stability_privamov_time}
    }
  \caption{Stability of the \FLI{} models on {\sc PrivaMov} \& {\sc Cabspotting} with $\epsilon=10^{-3}$.
  }
  \label{fig:stability}
\end{figure}

We further explore the capability of \FLI{} to capture stable models that group as many data samples as possible for the longest possible durations.
Fig.~\ref{fig:stability} reports on the time and the number of samples covered by the models of \FLI{} for the {\sc Cabspotting} and {\sc PrivaMov} datasets.
One can observe that the stability of \FLI{} depends on the density of the considered datasets.
While \FLI{} only captures at most $4$ samples for $90$\% of the models stored in {\sc Cabspotting} (Fig.~\ref{fig:stability_cabspotting_samples}), it reaches up to $2{,}841$ samples in the context of {\sc PrivaMov} (Fig.~\ref{fig:stability_privamov_samples}), which samples \gls{gps} locations at a higher frequency than {\sc Cabspotting}.
This is confirmed by Fig.~\ref{fig:stability_cabspotting_time} and~\ref{fig:stability_privamov_time}, which report a time coverage of $202$\,ms and $3{,}602$\,ms for $90$\% of \FLI{} models in {\sc Cabspotting} and {\sc PrivaMov}, respectively.
Given that {\sc PrivaMov} is a larger dataset than {\sc Cabspotting} ($7.2$\,GB vs. $388$\,MB), one can conclude that \FLI{} succeeds in scaling with the volume of data to be stored.

\subsection{Enabling \emph{in-situ} big data processing}

Location data is not only highly sensitive privacy-wise but also crucial for location-based services.
While \glspl{lppm} have been developed to protect user locations, they are generally used on the server where the data is aggregated.
The user location data is thus exposed to classical threats, such as malicious users, man in the middle, or database leaks.
To avoid such threats, one privacy-preserving solution would be to keep the data in the device where it is produced until it is sufficiently obfuscated to be shared with a third party.
With \gls{gps} data, this protection mechanism must be undertaken by a device-local \gls{lppm}.
Evaluating the privacy of the resulting trace must also be performed locally, by executing attacks on the obfuscated data.
Both processes require storing all the user mobility traces in the mobile device.

While existing approaches have simulated this approach~\citep{eden}, no real deployment has ever been reported.
In this section, we show that using \FLI{} enables overcoming one of the memory hurdles of constrained devices.
We use \FLI{} to store entire \gls{gps} traces in mobile devices, execute \acrshort{poi} attacks, and protect the traces using the \acrshort{lppm} {\sc Promesse}~\citep{promesse}.

{\sc Promesse}~\citep{promesse} is an \acrshort{lppm} that intends to hide \glspl{poi} from a mobility trace by introducing a negligible spatial error.
To do so, {\sc Promesse} smooths the trajectories by replacing the mobility trace with a new one applying a constant speed while keeping the same starting and ending timestamps.
The new trace $T'$ is characterized by the distance $\delta$ between two points.
First, additional locations are inserted by considering the existing locations one by one in chronological order.
If the distance between the last generated location $T'[i]$ and the current one $T[c]$ is below $\delta$, this location is discarded.
Otherwise, $T'[i+1]$ is not defined as the current location $T[c]$, but the location between $T'[i]$ and $T[c]$, such that the distance between $T'[i]$ and $T'[i+1]$ is equal to $\delta$.
Once all the locations included in the new mobility trace are defined, the timestamps are updated to ensure that the period between the two locations is the same, keeping the timestamps of the first and last locations unchanged.
The resulting mobility trace is protected against \acrshort{poi} attacks while providing high spatial accuracy.

\begin{figure}[t]
  \captionsetup[subfloat]{farskip=0pt}
  \center
  \includegraphics[width=.8\linewidth]{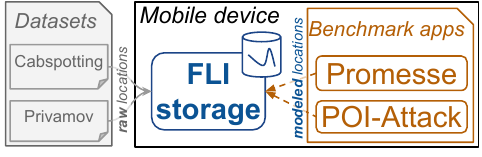}
  \caption{Architectural diagram of our \textit{in-situ} big data processing experiment on a mobile device.}
  \label{fig:fli_system}
\end{figure}

We, therefore, implemented a mobile version of \acrshort{poi} attacks and {\sc Promesse}~\citep{promesse} using the Flutter library of \FLI{} and evaluated its performances on the {\sc PrivaMov} dataset.
Fig.~\ref{fig:fli_system} depicts the resulting experimental deployments of \gls{poi} attacks and the corresponding \gls{lppm} in mobile devices, alongside \FLI.
Our experiments are performed using a publicly available application~\citep{insitulppm}.

\begin{table*}
  \centering
  \caption{
    \label{tab:poi_computation_times}
    Execution time for {\sc PrivaMov} {\tt user 1} on different mobile platforms. 
    Previous deployments of these algorithms were made on desktop computers. 
    \FLI now enables their instanciation on ubiquitous devices, with reasonable processing times.
  }

  \begin{tabular}{l c c c c c}
    \hline
    \textbf{\textsc{Task}} & \textbf{Moto~Z} & \textbf{Fair Phone 3} & \textbf{Google Pixel 7 Pro} & \textbf{iPhone 12} & \textbf{iPhone 14+}\\
    \hline
    \textsc{Promesse} & 1.5s & 1.3s & 0.4s & 0.2s & 0.2s\\
    \hline
    \textsc{POI-Attack} & 114.4s & 109.2s & 30s & 18.8s & 19.4s\\
    \hline
  \end{tabular}
\end{table*}

Given that {\sc PrivaMov} is the largest dataset, Table~\ref{tab:poi_computation_times} reports on the worst-case processing times observed for {\tt user 1} on different mobile platforms.
This captures a user mobility trace of $4,341,716$ locations reported by a mobile device.
It shows that applying \textsc{Promesse} to the {\tt user 1} mobility trace takes 1 second to run on all devices, on average.
More interestingly, \textsc{POI-Attack} takes no longer than 30 seconds ($\pm~43ms$) to run on the latest generation hardware, both on Android and iOS.
This is a significant improvement compared to the state-of-the-art implementation~\citep{primault2014differentially}, which requires one hour to run on a desktop computer.
We believe that this is a critical step forward towards improving user privacy as all \gls{lppm} experiments until today were either simulated or centralized, contrary to what the literature suggests~\citep{sprinkler}.

\begin{figure}
  \center
  \includegraphics[width=.6\linewidth]{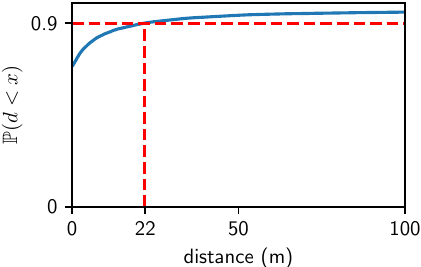}
  \caption{
    Distances distribution when using \stay{} on {\sc Cabspotting}.
    The distances between the \glspl{poi} are obtained using \stay{} and their closest counterparts, obtained with the traditional \gls{poi} attack.
    Except for a few extreme values, the values are close: more than $68\%$ are the same and $90$\% of the \glspl{poi} are at a distance lower than $22$ meters than a "real" one.
  }
  \label{fig:xp_distances_poi}
\end{figure}

\begin{figure}
  \centering
  \includegraphics[scale=0.40]{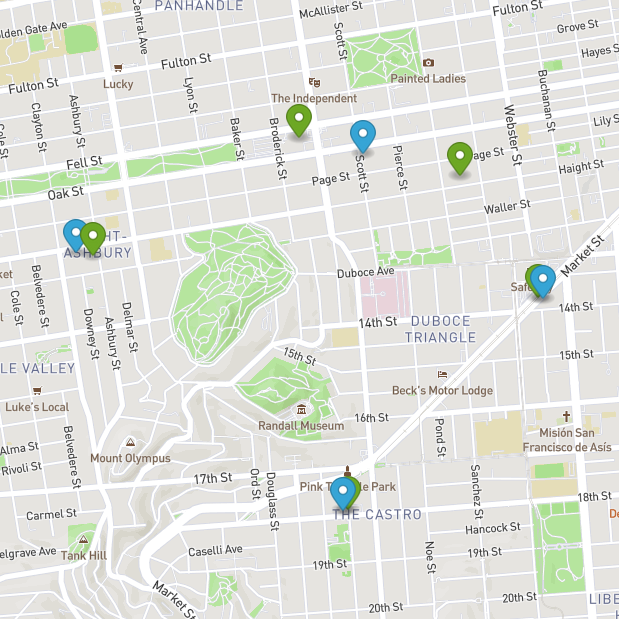}
  \caption{Example of \acrshort{poi} of {\tt user 0} from {\sc Cabspotting} using the standard \acrshort{poi} algorithm on raw data (in {\color{blue}blue}) and those obtained on the data modeled by \FLI{} and computed with \STAY (in {\color{green}green}).
  The inferred \acrshort{poi} are very close on average, but there are some outliers.
  }
  \label{image:poi-cabspotting-user-0}
\end{figure}

In addition to speed, the quality of the inferred \glspl{poi} is the most salient concern about \stay.
We assess the quality by computing the distances to the \glspl{poi} obtained from the \gls{poi}-attack on {\sc Cabspotting}.
We choose {\sc Cabspotting} because computing it on {\sc PrivaMov} is prohibitive in terms of computation time.
Fig.~\ref{fig:xp_distances_poi} depicts the distribution of the distances below $100$ meters: more than $68\%$ are the same and $90\%$ of the \glspl{poi} are at a distance lower than $22$ meters from actual ones.
Fig.~\ref{image:poi-cabspotting-user-0} shows some of the \acrshort{poi} inferred from raw samples and the standard \acrshort{poi}-attack (in blue) and \acrshort{poi} obtained with \FLI{} and \STAY: they are different but very close,
highlighting the negligible impact for applications that depend on location data.

Finally, Table~\ref{tab:poi_effect_algo} highlights that {\sc Promesse} works well independently of the use of \FLI{} or \STAY: it successfully hides all the \acrshort{poi}.
Therefore, \stay{} provides an important speed-up without altering the quality of \glspl{poi}.
Note that \FLI{} was not used in this case, as the performances of \stay{} are orthogonal to the use of a temporal database to model the samples.

\begin{table*}
  \centering
  \caption{
    \label{tab:poi_effect_algo}
    Impact of \FLI and \STAY on the number of inferred \glspl{poi} from {\tt user 0} trace in {\sc Cabspotting}.
    Thanks to \FLI and \STAY, {\sc Promesse} succeeds to protect user privacy at the edge.
  }
  \resizebox{0.7\linewidth}{!}{
    \begin{tabular}{c c >{\columncolor{Gray}}c c >{\columncolor{Gray}}c}
      & \multicolumn{2}{c}{without {\sc Promesse}} & \multicolumn{2}{c}{with {\sc Promesse}} \\[-.2em]
      & \multicolumn{2}{c}{$\overbrace{\hspace{10em}}$} & \multicolumn{2}{c}{$\overbrace{\hspace{10em}}$} \\[-.5em]
      \bf Algorithm & \bf Raw \glspl{poi} & \bf \FLI & \bf Raw \glspl{poi} & \bf \FLI \\ \hline
      \gls{poi}-attack  & 30 & 31 & 0 & 0 \\
      \STAY & 30 & 30 & 0 & 0 \\
      \hline
      \gls{poi}-attack $\cap$ \STAY & 21 & 20 & - & -\\
      \hline
    \end{tabular}
  }
\end{table*}

\subsection{Beyond Location Streams}\label{ssec:beyond_location_streams}
In this paper, \FLI{} was mainly tested against location streams, but our proposal efficiently approximates any type of signal that varies mostly linearly: timestamps, accelerations, temperature, pressure, humidity, light, proximity, air quality, etc.
This makes \FLI{} a valuable candidate in ubiquitous contexts, as many physical quantities captured by e.g. IoT sensors have a piecewise linear behavior.
To showcase different scenarios, we benchmark the storage of timestamps, device accelerations and heartbeat data using \FLI.
In any case, we use our $\epsilon$-tuning script to capture the most appropriate value to store the data with a reduced error.

\paragraph{Storing timestamps.}
In all the previous experiments, the timestamps were not modeled by \FLI, as we expect the user to query the time at which she is interested in the samples.
However, it is straightforward to store irregular timestamps using \FLI: we store couples $(i,t_i)$ with $t_i$ being the $i^{\text{th}}$ inserted timestamp.
The nature of the timestamps makes them a good candidate for modeling, as insertion rates are generally fixed, or vary linearly.
To assess the efficiency of \FLI{} for storing timestamps, we stored all the timestamps of the {\tt user 1} of the {\sc PrivaMov} dataset with $\epsilon=1$---\emph{i.e.}, we tolerate an error of one second per estimate.
The $4{,}341{,}716$ timestamps were stored using $26{,}862$ models for a total of $80{,}592$ floats and an overall gain of $98\%$, with a \emph{mean average error} (MAE) of $0.246$ second.
Hence, not only does the use of \FLI{} result in drastic memory savings, but it also provides accurate estimations.

\paragraph{Storing device accelerations.}
Accelerometer data is important for many context-aware applications, including transport mode detection~\citep{yu2014big,fang2016transportation,wang2019enabling}.
Coupled with \gls{gps} information, it is possible to infer whether the user is walking, biking, taking a car or a tram.
However, storing the output of a mobile accelerometer is particularly challenging, as it generates hundreds of noisy 3D samples per second.
Our implementation is publicly available~\citep{flairaccelerometerexampleapp}.

We store $10{,}000$ consecutive accelerometer samples with \FLI{} and, for every $100$ insertions, we report on the size of the file and the relative gain.
We use \FLI{} with $\epsilon=1$ as the accelerometer has high variability, even when the mobile is stationary.
Fig.~\ref{fig:xp_accelerometer} reports on a constant memory footprint of \FLI{} during the experiment, while providing a high-level accuracy.

\renewcommand{\scaleBenchmarking}{0.58}
\begin{figure}
 \captionsetup[subfloat]{farskip=0pt}  \center
 \subfloat[Model size]{
   \includegraphics[scale=\scaleBenchmarking]{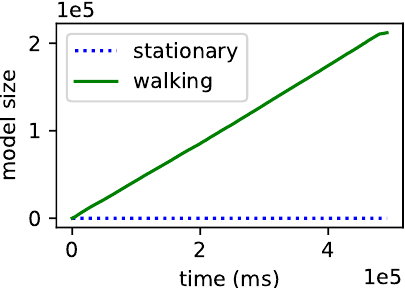}
   \label{fig:xp_accelerometer_size}
 }
 \subfloat[Memory gain]{
   \includegraphics[scale=\scaleBenchmarking]{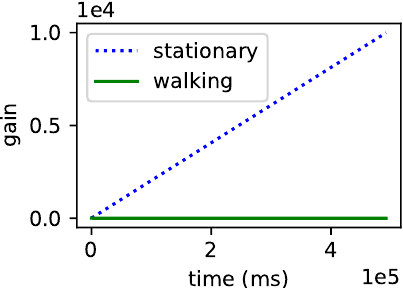}
   \label{fig:xp_accelerometer_gain}
 }
 \caption{
   Model size and gain of \FLI{} storing accelerometer values ($\epsilon=1$).
   \FLI{} reports a constant memory whenever stationary, and a small gain ($>\times 1.39$) when walking.
 }
 \label{fig:xp_accelerometer}
\end{figure}

\paragraph{Storing heartbeat pulses.}
We downloaded pulse-to-pulse intervals, which oscillate between $500$ and $1,100$\,ms and are reported as the time duration between cardiac pulses to the
timestamp of the original pulse, from a \textit{Polar Ignite 2}~\citep{ignite} smartwatch, gathering $28{,}294{,}762$ samples covering the 12 months of 2023, as a file of $259.1$\,MB.
Using \FLI{} to model this dataset with an accuracy of $\epsilon=100$ reports on a non-negligible storage space gain of $26.44\%$, with a \emph{mean error} of $22.74$ milliseconds.
\FLI{} is thus a suitable solution to store data streams produced by various sensors of wearable and mobile devices, which could find application in e.g. human context recognition~\citep{detailedHumanContext2017}.\\

To conclude, our implementation of the data stream storage solution, \FLI, enables the effective deployment of more advanced techniques, such as EDEN~\citep{eden} or HMC~\citep{maouche2018hmc}.
This may require new algorithms, such as \stay{}, but it enables {\it in~situ} data privacy protection before sharing any sensitive information.
We believe that this is a critical step forward towards improving user privacy as all \gls{lppm} experiments until today were either centralized or simulated.

Thanks to its \FLI{} storage layer and \stay{} attack layer, \INTACT improves privacy, and the cost of this improved privacy is \textit{in situ} data computation, which requires running potentially heavy data tasks on constrainted devices.
We report on the execution time of \textsc{POI-Attack} in Table~\ref{tab:poi_computation_times}, and argue computation time is the price to pay for an increased privacy.

  \section{Threats to Validity}
\label{sec:threats}
While the combination of \FLI and \STAY succeeds in embedding \glspl{lppm} within mobile devices and increasing user privacy, our results might be threatened by some variables we considered.

The hardware threats relate to the classes of constrained devices we considered.
In particular, we focused on the specific case of smartphones, which is the most commonly deployed mobile device in the wild.
To limit the bias introduced by a given hardware configuration, we deployed \FLI on both recent Android and iOS smartphones for most of the reported experiments, while we also considered the impact of hardware configurations on the reported performances.

Another potential bias relates to the mobility datasets we considered in the context of this paper.
To limit this threat, we evaluated our solutions on two established mobility datasets, {\sc Cabspotting} and {\sc PrivaMov}, which exhibit different characteristics.
Yet, we could further explore the impact of these characteristics (sampling frequency, number of participants, duration and scales of the mobility traces).
Beyond mobility datasets, we could %
consider the evaluation of other \acrshort{iot} data streams, such as air quality metrics, to assess the capability of \FLI to handle a wide diversity of data streams.
To mitigate this threat, we reported on the storage of timestamps, acceleration and heartbeat data in addition to 2-dimensional locations.

Although \FLI increases storage capacity through data modeling, it might still reach the storage limit of its host device if using a constant $\epsilon$ parameter (which drives the compression rate).
To address this issue, we could dynamically adapt data compression to fit a storage size constraint.
Toward this end, An~\emph{et~al.}~\citep{yanzhe2022tvstore} propose an interesting time-aware adaptive compression rate, based on the claim that data importance varies with its age.

Beyond the current implementation reported in this article, one could envision a native integration of \INTACT in
the Android and iOS operating systems to enable \glspl{lppm} for any legacy application.
This technical challenge mostly consists of packaging \FLI and \STAY as a new \texttt{LocationProvider} in
Android~\citep{locationproviderandroid} and a new \texttt{CLLocationManager} in iOS~\citep{locationproviderios}.
Interestingly, such a native integration of \INTACT can allow end users to configure the list of enabled
\glspl{lppm} and their related settings through the operating system control panel.
Although not implemented in practice, \INTACT includes a GPS provider part, made of \acrshort{lppm} protections
and \acrshort{poi} attacks, which can supply obfuscated locations to LBS apps;
this is however straightforward to implement on mobile devices, using system abstractions both available in
Android~\citep{locationproviderandroid} and iOS~\citep{locationproviderios}\@.

The increased storage capacity offered by \FLI not only allows for unlimited mobility data storage, but also allows
applying \textit{in-situ} \acrshort{lppm}s requiring lots of data to work, for instance, those offering \textit{k} and
\textit{l}-anonymity guarantees by hiding user among others~\citep{sweeney2002k,machanavajjhala2007diversity}.

Our implementations of \FLI and \STAY
may suffer from software bugs that affect the reported performances.
To limit this threat, we make the code of our libraries and applications freely available to encourage the
reproducibility of our results and share the implementation decisions we took as part of the current implementation.

Finally, our results might strongly depend on the parameters we pick to evaluate our contributions.
While \FLI performances (gain, memory footprint) vary depending on the value of the $\epsilon$ parameter, we considered
a sensitive analysis of this parameter and we propose a default value $\epsilon=10^{-3}$ that delivers a minimum memory
gain that limits the modeling error.

\section{Perspectives}
\label{sec:discussion}
Beyond the deployment of \acrshort{lppm} at the edge of the network, we believe that our contributions open new avenues
for improving user privacy and more generally enabling data processing at the edge.
This includes the implementation of alternative \acrshort{lppm}s, the implementation of differential privacy algorithms,
like \emph{k-anonymity} and \emph{l-diversity}, but also the inclusion of more privacy attacks to provide users with
privacy reports for the data they intend to share. %

\FLI demonstrates that increasing the storage capacity of data streams on constrained devices offers new distributed
computation models, inspired by the big data principles, that can deploy automated processing tasks on edge devices.
Beyond \acrshort{lppm}s, one can think about the involvement of mobile devices in a federated or decentralized learning
environment to train models without compromising sensitive personal information.
Furthermore, we demonstrated with \STAY that one can execute privacy attacks at the scale of a device, hence delivering
support to assess the sensibility of a trained model before sharing it with some third party.

Regarding \FLI storage capabilities, we think that the integration of compressed data could contribute to the further
reduction of the memory footprint of the data streams stored on mobile devices, %
enabling computations over the older history of streams.

Regarding the $\epsilon$ parameter, we believe that its choice could not only be further automated by systematically applying the selection routine (outlined in Sec.~\ref{ssec:parameter_tuning}) before the modeling starts, but could even be dynamically updated.
This would allow for a dynamic data compression rate, which could be driven for instance by the data's age or its asserted importance. 
We plan on tackling this issue in future works.

While the literature in the field of mobile privacy is rich, we hope open source implementation of our Flutter libraries will %
encourage the research and developer communities to embed our contributions to strengthen the privacy of their users.
We believe that \FLI and \STAY are key software assets that can stimulate the mobile database and privacy communities
to deliver effective solutions on mobile devices, which are widely deployed nowadays.

\section{Conclusion}\label{sec:conclusion}

Mobile devices are incredible producers of data streams, which are often forwarded to remote third-party services for
storage and processing.
This data processing pattern might be the source of privacy breaches, as the raw data may leak sensitive personal information.
Furthermore, the volume of data to be processed may require huge storage capacity, from mobile devices to remote servers,
and network capacity to deal with the increasing number of devices deployed in the wild.

On the other hand, %
while \glspl{lppm} are a promising solution to protect user locations, they are not widely deployed in practice, mostly because
of their high computational cost, which is prohibitive on mobile devices.

We, therefore, proposed \INTACT, a software framework that enables \glspl{lppm} on mobile devices by leveraging the increased storage capacity of mobile devices.

The contributions of \INTACT are threefold: we leveraged {\em i)} a compact storage system based on a piece-wise linear
model dubbed \FLI (proposed in~\citep{fli}), {\em ii)} introduced a new way to compute \glspl{poi}, called \stay, and finally
{\em iii)} demonstrated how \FLI could unlock device-local privacy protections on time series while using machine learning.

Additionally, we provided an operational Flutter package implementing them along with other existing temporal databases.

As a matter of perspectives, we therefore believe that \FLI can be used as a building block for the development of privacy-friendly
mobile applications, which can store and process data locally, without the need to rely on remote third-party services.
In particular, one can explore the deployment of federated learning algorithms on mobile devices, which can be used to train machine
learning models without compromising sensitive personal information.
On the longer term, \FLI paves the way to implement decentralized machine learning algorithms in mobile devices by leveraging
device-to-device communication models, like {\sc Sprinkler}~\citep{sprinkler}.
For example, we believe \acrfull{spi} captured by ubiquitous devices should be anonymized locally {\it before} any data exchange.

While \FLI can store tremendous data on mobile devices, \stay provides an important speed-up to reduce the total computation time
of \gls{poi} attacks by several orders of magnitude, making them suitable for mobile computing.

By sharing this \INTACT framework with mobile developers, our contribution is an important step forward towards the real deployment
of \glspl{lppm} and, more generally, privacy-friendly data-intensive workloads at the edge (\emph{e.g.}, federated learning on mobile
phones).\\

\begin{acknowledgements}
This research was supported in part by the Groupe La Poste, sponsor of the Inria Foundation, in the framework of the FedMalin Inria Challenge.
\end{acknowledgements}

\begin{contributions}
Rémy Raes and Olivier Ruas contributed to the conception of the study and performed experiments.
Both are the main contributors and writers of this manuscript.
Adrien Luxey-Bitri and Romain Rouvoy supervised the work, wrote and revised the text.
All authors read and approved the final manuscript.
This work is an extension of \cite{fli}.
\end{contributions}

\begin{interests}
The authors declare no conflict of interest.
\end{interests}

\begin{materials}
Source code of the \INTACT framework is publicly available in the Software Heritage repository at:
\url{https://archive.softwareheritage.org/browse/origin/directory/?branch=refs/tags/jisa-2024-artefacts&origin_url=https://gitlab.inria.fr/Spirals/temporaldb_apps.git}
\end{materials}

  \bibliographystyle{apalike-sol}
  \bibliography{references}

\begin{thebibliography}{}

\bibitem[An {\em et~al}., 2022]{yanzhe2022tvstore}
An, Y., Su, Y., Zhu, Y., and Wang, J. (2022).
\newblock {TVStore}: Automatically bounding time series storage via
  {Time-Varying} compression.
\newblock In {\em 20th USENIX Conference on File and Storage Technologies (FAST
  22)}, pages 83--100, Santa Clara, CA. USENIX Association.

\bibitem[Andr{\'e}s {\em et~al}., 2013]{geoI}
Andr{\'e}s, M.~E., Bordenabe, N.~E., Chatzikokolakis, K., and Palamidessi, C.
  (2013).
\newblock Geo-indistinguishability: Differential privacy for location-based
  systems.
\newblock In {\em Proceedings of the 2013 ACM SIGSAC conference on Computer \&
  communications security}, pages 901--914.

\bibitem[Apple, 2013]{locationproviderios}
Apple (2013).
\newblock i{OS} \texttt{CLLocationManager} documentation.
\newblock
  \url{https://developer.apple.com/documentation/corelocation/cllocationmanager}.
\newblock Last accessed on April 21st, 2024.

\bibitem[Bellet {\em et~al}., 2017]{bellet2017fast}
Bellet, A., Guerraoui, R., Taziki, M., and Tommasi, M. (2017).
\newblock {\em Fast and differentially private algorithms for decentralized
  collaborative machine learning}.
\newblock PhD thesis, INRIA Lille.

\bibitem[Berlin and Van~Laerhoven, 2010]{emswab}
Berlin, E. and Van~Laerhoven, K. (2010).
\newblock An on-line piecewise linear approximation technique for wireless
  sensor networks.
\newblock In {\em IEEE Local Computer Network Conference}, pages 905--912.
  IEEE.

\bibitem[Binder, 2019]{drift}
Binder, S. (2019).
\newblock Drift library.
\newblock \url{https://pub.dev/packages/drift}.
\newblock Last accessed on April 21st, 2024.

\bibitem[Blalock {\em et~al}., 2018]{blalockSprintzTimeSeries2018}
Blalock, D., Madden, S., and Guttag, J. (2018).
\newblock Sprintz: {{Time Series Compression}} for the {{Internet}} of
  {{Things}}.
\newblock {\em Proceedings of the ACM on Interactive, Mobile, Wearable and
  Ubiquitous Technologies}, 2(3). DOI: 10.1145/3264903.

\bibitem[Cerf {\em et~al}., 2017]{pulp}
Cerf, S., Primault, V., Boutet, A., Mokhtar, S.~B., Birke, R., Bouchenak, S.,
  Chen, L.~Y., Marchand, N., and Robu, B. (2017).
\newblock {PULP:} achieving privacy and utility trade-off in user mobility
  data.
\newblock In {\em 36th {IEEE} Symposium on Reliable Distributed Systems, {SRDS}
  2017, Hong Kong, September 26-29, 2017}, pages 164--173. {IEEE} Computer
  Society. DOI: 10.1109/SRDS.2017.25.

\bibitem[Dollinger and Junginger, 2014]{objectbox}
Dollinger, V. and Junginger, M. (2014).
\newblock Objectbox database.
\newblock \url{https://objectbox.io}.
\newblock Last accessed on April 21st, 2024.

\bibitem[Dwork, 2008]{differential}
Dwork, C. (2008).
\newblock Differential privacy: A survey of results.
\newblock In {\em International conference on theory and applications of models
  of computation}, pages 1--19. Springer.

\bibitem[Fang {\em et~al}., 2016]{fang2016transportation}
Fang, S.-H., Liao, H.-H., Fei, Y.-X., Chen, K.-H., Huang, J.-W., Lu, Y.-D., and
  Tsao, Y. (2016).
\newblock Transportation modes classification using sensors on smartphones.
\newblock {\em Sensors}, 16(8):1324.

\bibitem[Galakatos {\em et~al}., 2019]{galakatos2019fiting}
Galakatos, A., Markovitch, M., Binnig, C., Fonseca, R., and Kraska, T. (2019).
\newblock {FIT}ing-tree: A data-aware index structure.
\newblock In {\em Proceedings of the 2019 International Conference on
  Management of Data}, pages 1189--1206.

\bibitem[Gambs {\em et~al}., 2014]{gambs2014anonymization}
Gambs, S., Killijian, M.-O., and del Prado~Cortez, M.~N. (2014).
\newblock De-anonymization attack on geolocated data.
\newblock {\em Journal of Computer and System Sciences}, 80(8):1597--1614.

\bibitem[Google, 2011]{locationproviderandroid}
Google (2011).
\newblock Android \texttt{LocationManager} documentation.
\newblock
  \url{https://developer.android.com/reference/android/location/LocationManager#addTestProvider(java.lang.String,%20android.location.provider.ProviderProperties,%20java.util.Set%3Cjava.lang.String%3E)}.
\newblock Last accessed on April 21st, 2024.

\bibitem[Google, 2018]{flutterframework}
Google (2018).
\newblock Flutter framework.
\newblock \url{https://flutter.dev/}.
\newblock Last accessed on April 21st, 2024.

\bibitem[Gr{\"u}tzmacher {\em et~al}., 2018]{grutzmacher2018time}
Gr{\"u}tzmacher, F., Beichler, B., Hein, A., Kirste, T., and Haubelt, C.
  (2018).
\newblock Time and memory efficient online piecewise linear approximation of
  sensor signals.
\newblock {\em Sensors}, 18(6):1672.

\bibitem[Hariharan and Toyama, 2004]{hariharan2004project}
Hariharan, R. and Toyama, K. (2004).
\newblock Project lachesis: parsing and modeling location histories.
\newblock In {\em International Conference on Geographic Information Science},
  pages 106--124. Springer.

\bibitem[InfluxData, 2013]{InfluxDB}
InfluxData (2013).
\newblock Influx{DB}.
\newblock \url{https://www.influxdata.com/products/influxdb-overview/}.
\newblock Last accessed April 21st, 2024.

\bibitem[Keogh {\em et~al}., 2001]{swab}
Keogh, E., Chu, S., Hart, D., and Pazzani, M. (2001).
\newblock An online algorithm for segmenting time series.
\newblock In {\em Proceedings 2001 IEEE international conference on data
  mining}, pages 289--296. IEEE.

\bibitem[Khalfoun {\em et~al}., 2021]{eden}
Khalfoun, B., Ben~Mokhtar, S., Bouchenak, S., and Nitu, V. (2021).
\newblock {EDEN}: Enforcing location privacy through re-identification risk
  assessment: A federated learning approach.
\newblock {\em Proceedings of the ACM on Interactive, Mobile, Wearable and
  Ubiquitous Technologies}, 5(2). DOI: 10.1145/3463502.

\bibitem[Liu {\em et~al}., 2008]{liu2008novel}
Liu, X., Lin, Z., and Wang, H. (2008).
\newblock Novel online methods for time series segmentation.
\newblock {\em IEEE Transactions on Knowledge and Data Engineering},
  20(12):1616--1626.

\bibitem[Luxey {\em et~al}., 2018]{sprinkler}
Luxey, A., Bromberg, Y.-D., Costa, F.~M., Lima, V., da~Rocha, R. C.~A., and
  Taïani, F. (2018).
\newblock Sprinkler: A probabilistic dissemination protocol to provide fluid
  user interaction in multi-device ecosystems.
\newblock In {\em IEEE International Conference on Pervasive Computing and
  Communications (PerCom)}, pages 1--10.

\bibitem[Machanavajjhala {\em et~al}., 2007]{machanavajjhala2007diversity}
Machanavajjhala, A., Kifer, D., Gehrke, J., and Venkitasubramaniam, M. (2007).
\newblock l-{D}iversity: Privacy beyond k-anonymity.
\newblock {\em ACM Transactions on Knowledge Discovery from Data (TKDD)},
  1(1):3--es.

\bibitem[Maouche {\em et~al}., 2018]{maouche2018hmc}
Maouche, M., Ben~Mokhtar, S., and Bouchenak, S. (2018).
\newblock {HMC}: Robust privacy protection of mobility data against multiple
  re-identification attacks.
\newblock {\em Proceedings of the ACM on Interactive, Mobile, Wearable and
  Ubiquitous Technologies}, 2(3):1--25.

\bibitem[Maouche {\em et~al}., 2017]{maouche2017ap}
Maouche, M., Mokhtar, S.~B., and Bouchenak, S. (2017).
\newblock Ap-attack: a novel user re-identification attack on mobility
  datasets.
\newblock In {\em Proceedings of the 14th EAI International Conference on
  Mobile and Ubiquitous Systems: Computing, Networking and Services}, pages
  48--57.

\bibitem[Meftah {\em et~al}., 2019]{meftah2019fougere}
Meftah, L., Rouvoy, R., and Chrisment, I. (2019).
\newblock Fougere: user-centric location privacy in mobile crowdsourcing apps.
\newblock In {\em IFIP International Conference on Distributed Applications and
  Interoperable Systems}, pages 116--132. Springer.

\bibitem[Moawad {\em et~al}., 2015]{greycat}
Moawad, A., Hartmann, T., Fouquet, F., Nain, G., Klein, J., and Le~Traon, Y.
  (2015).
\newblock Beyond discrete modeling: A continuous and efficient model for
  {I}o{T}.
\newblock In {\em 2015 ACM/IEEE 18th International Conference on Model Driven
  Engineering Languages and Systems (MODELS)}, pages 90--99. IEEE.

\bibitem[Mokhtar {\em et~al}., 2017]{privamov}
Mokhtar, S.~B., Boutet, A., Bouzouina, L., Bonnel, P., Brette, O., Brunie, L.,
  Cunche, M., D'Alu, S., Primault, V., Raveneau, P., {\em et~al}. (2017).
\newblock {PRIVA'MOV}: Analysing human mobility through multi-sensor datasets.
\newblock In {\em NetMob 2017}.

\bibitem[Piorkowski {\em et~al}., 2009]{piorkowski2009crawdad}
Piorkowski, M., Sarafijanovic-Djukic, N., and Grossglauser, M. (2009).
\newblock {CRAWDAD} data set epfl/mobility (v. 2009-02-24).

\bibitem[Polar, 2021]{ignite}
Polar (2021).
\newblock Ignite 2.
\newblock \url{https://www.polar.com/en/ignite2}.
\newblock Last accessed on April 21st, 2024.

\bibitem[Primault {\em et~al}., 2014]{primault2014differentially}
Primault, V., Mokhtar, S.~B., Lauradoux, C., and Brunie, L. (2014).
\newblock Differentially private location privacy in practice.
\newblock {\em arXiv preprint arXiv:1410.7744}.

\bibitem[Primault {\em et~al}., 2015]{promesse}
Primault, V., Mokhtar, S.~B., Lauradoux, C., and Brunie, L. (2015).
\newblock Time distortion anonymization for the publication of mobility data
  with high utility.
\newblock In {\em 2015 IEEE Trustcom/BigDataSE/ISPA}, volume~1, pages 539--546.
  IEEE.

\bibitem[Raes {\em et~al}., 2022a]{flairaccelerometerexampleapp}
Raes, R., Ruas, O., Luxey-Bitri, A., and Rouvoy, R. (2022a).
\newblock \fli accelerometer example application.
\newblock
  \href{https://archive.softwareheritage.org/browse/revision/570c744aaa82ce8c8f75fce53234013b001872fb/?origin_url=https://gitlab.inria.fr/Spirals/temporaldb_apps.git&path=temporaldb/example&revision=570c744aaa82ce8c8f75fce53234013b001872fb}{Hosted
  on Software Heritage}.
\newblock Last accessed on November 4th, 2024.

\bibitem[Raes {\em et~al}., 2022b]{flairimplementation}
Raes, R., Ruas, O., Luxey-Bitri, A., and Rouvoy, R. (2022b).
\newblock \fli implementation.
\newblock
  \href{https://archive.softwareheritage.org/browse/revision/570c744aaa82ce8c8f75fce53234013b001872fb/?origin_url=https://gitlab.inria.fr/Spirals/temporaldb_apps.git&path=temporaldb&revision=570c744aaa82ce8c8f75fce53234013b001872fb}{Hosted
  on Software Heritage}.
\newblock Last accessed on November 4th, 2024.

\bibitem[Raes {\em et~al}., 2022c]{insitulppm}
Raes, R., Ruas, O., Luxey-Bitri, A., and Rouvoy, R. (2022c).
\newblock In-situ \uppercase{LPPM}.
\newblock
  \href{https://archive.softwareheritage.org/browse/revision/570c744aaa82ce8c8f75fce53234013b001872fb/?origin_url=https://gitlab.inria.fr/Spirals/temporaldb_apps.git&path=in_situ_lppm&revision=570c744aaa82ce8c8f75fce53234013b001872fb}{Hosted
  on Software Heritage}.
\newblock Last accessed on November 4th, 2024.

\bibitem[Raes {\em et~al}., 2022d]{benchmarkingmemoryspace}
Raes, R., Ruas, O., Luxey-Bitri, A., and Rouvoy, R. (2022d).
\newblock Memory space benchmarking application.
\newblock
  \href{https://archive.softwareheritage.org/browse/revision/570c744aaa82ce8c8f75fce53234013b001872fb/?origin_url=https://gitlab.inria.fr/Spirals/temporaldb_apps.git&path=benchmarking_memory_space&revision=570c744aaa82ce8c8f75fce53234013b001872fb}{Hosted
  on Software Heritage}.
\newblock Last accessed on November 4th, 2024.

\bibitem[Raes {\em et~al}., 2022e]{benchmarkingthroughput}
Raes, R., Ruas, O., Luxey-Bitri, A., and Rouvoy, R. (2022e).
\newblock Throughput benchmarking application.
\newblock
  \href{https://archive.softwareheritage.org/browse/revision/570c744aaa82ce8c8f75fce53234013b001872fb/?origin_url=https://gitlab.inria.fr/Spirals/temporaldb_apps.git&path=benchmarking_throughput&revision=570c744aaa82ce8c8f75fce53234013b001872fb}{Hosted
  on Software Heritage}.
\newblock Last accessed on November 4th, 2024.

\bibitem[Raes {\em et~al}., 2024]{fli}
Raes, R., Ruas, O., Luxey-Bitri, A., and Rouvoy, R. (2024).
\newblock {Compact Storage of Data Streams in Mobile Devices}.
\newblock In {\em {DAIS'24 - 24th International Conference on Distributed
  Applications and Interoperable Systems}}, Proceedings of the 24th
  International Conference on Distributed Applications and Interoperable
  Systems (DAIS'24), Groningen, Netherlands. {LNCS}.

\bibitem[Sweeney, 2002]{sweeney2002k}
Sweeney, L. (2002).
\newblock k-anonymity: A model for protecting privacy.
\newblock {\em International Journal of Uncertainty, Fuzziness and
  Knowledge-Based Systems}, 10(05):557--570.

\bibitem[Tamplin and Lee, 2012]{firebase}
Tamplin, J. and Lee, A. (2012).
\newblock Firebase services.
\newblock \url{https://firebase.google.com}.
\newblock Last accessed on April 21st, 2024.

\bibitem[{Timescale}, 2019]{timescale_building_2019}
{Timescale} (2019).
\newblock Building a distributed time-series database on {{PostgreSQL}}.
\newblock Last accessed on May 12th 2023.

\bibitem[Timescale{ }Inc, 2018]{timescale}
Timescale{ }Inc (2018).
\newblock Timescale database.
\newblock \url{https://www.timescale.com}.
\newblock Last accessed on April 21st, 2024.

\bibitem[Vaizman {\em et~al}., 2017]{detailedHumanContext2017}
Vaizman, Y., Ellis, K., and Lanckriet, G. (2017).
\newblock Recognizing {{Detailed Human Context}} in the {{Wild}} from
  {{Smartphones}} and {{Smartwatches}}.
\newblock {\em IEEE Pervasive Computing}, 16(4). DOI:
  10.1109/MPRV.2017.3971131.

\bibitem[Wang {\em et~al}., 2019]{wang2019enabling}
Wang, L., Gjoreski, H., Ciliberto, M., Mekki, S., Valentin, S., and Roggen, D.
  (2019).
\newblock Enabling reproducible research in sensor-based transportation mode
  recognition with the sussex-huawei dataset.
\newblock {\em IEEE Access}, 7:10870--10891.

\bibitem[Wolfson {\em et~al}., 1998]{ldr}
Wolfson, O., Chamberlain, S., Dao, S., Jiang, L., and Mendez, G. (1998).
\newblock Cost and imprecision in modeling the position of moving objects.
\newblock In {\em Proceedings 14th International Conference on Data
  Engineering}, pages 588--596. DOI: 10.1109/ICDE.1998.655822.

\bibitem[Xu {\em et~al}., 2015]{xu2015privacy}
Xu, K., Yue, H., Guo, L., Guo, Y., and Fang, Y. (2015).
\newblock Privacy-preserving machine learning algorithms for big data systems.
\newblock In {\em 2015 IEEE 35th international conference on distributed
  computing systems}, pages 318--327. IEEE.

\bibitem[y~Arcas, 2018]{y2018decentralized}
y~Arcas, B.~A. (2018).
\newblock Decentralized machine learning.
\newblock In {\em 2018 IEEE International Conference on Big Data (Big Data)},
  pages 1--1. IEEE.

\bibitem[Yu {\em et~al}., 2014]{yu2014big}
Yu, M.-C., Yu, T., Wang, S.-C., Lin, C.-J., and Chang, E.~Y. (2014).
\newblock Big data small footprint: The design of a low-power classifier for
  detecting transportation modes.
\newblock {\em Proceedings of the VLDB Endowment}, 7(13):1429--1440.

\bibitem[Zhou {\em et~al}., 2004]{zhou2004discovering}
Zhou, C., Frankowski, D., Ludford, P., Shekhar, S., and Terveen, L. (2004).
\newblock Discovering personal gazetteers: an interactive clustering approach.
\newblock In {\em Proceedings of the 12th annual ACM international workshop on
  Geographic information systems}, pages 266--273.

\end{thebibliography}

\end{document}